\newcommand{\diracslash}[1]{#1\llap{/\kern2pt}}

\newcommand{\be}{\begin{equation}}
\newcommand{\ee}{\end{equation}}
\newcommand{\bea}{\begin{eqnarray}}
\newcommand{\eea}{\end{eqnarray}}
\newcommand{\ba}[1]{\begin{array}{#1}}
\newcommand{\ea}{\end{array}}

\newcommand{\bt}{\begin{tabular}}
\newcommand{\et}{\end{tabular}}
\newcommand{\Tr}{{\rm Tr}}

\newcommand{\beas}{\begin{eqnarray*}}
\newcommand{\eeas}{\end{eqnarray*}}

\documentclass[prd,aps,floats,nofootinbib,showpacs,floatfix]{revtex4}
\usepackage{graphicx}
\begin{document}

\title{D-mesons and charmonium states in hot isospin asymmetric 
strange hadronic matter }
\author{Arvind Kumar}
\email{iitd.arvind@gmail.com}
\affiliation{Department of Physics, Indian Institute of Technology, Delhi,
Hauz Khas, New Delhi -- 110 016, India}
\author{Amruta Mishra}
\email{amruta@physics.iitd.ac.in,mishra@th.physik.uni-frankfurt.de}
\affiliation{Department of Physics, Indian Institute of Technology, Delhi,
Hauz Khas, New Delhi -- 110 016, India}

\def\be{\begin{equation}}
\def\ee{\end{equation}}
\def\bearr{\begin{eqnarray}}
\def\eearr{\end{eqnarray}}
\def\zbf#1{{\bf {#1}}}
\def\bfm#1{\mbox{\boldmath $#1$}}
\def\hf{\frac{1}{2}}
\def\kp{\zbf k+\frac{\zbf q}{2}}
\def\km{-\zbf k+\frac{\zbf q}{2}}
\def\hwo{\hat\omega_1}
\def\hwt{\hat\omega_2}
\begin{abstract}
We study the properties of $D$ and $\bar{D}$ mesons in hot isospin
asymmetric strange hadronic matter, arising due to their interactions with 
the hadrons in the hyperonic medium. The interactions of $D$ and $\bar{D}$ 
mesons with these light hadrons are derived by generalizing the chiral SU(3) 
model used for the study of hyperonic matter to SU(4). The nucleons, 
hyperons, the scalar isoscalar meson, $\sigma$ and the scalar-isovector 
meson, $\delta$ as modified in the strange hadronic matter, modify the 
masses of $D$ and $\bar{D}$ mesons. It is found that as compared to 
the $\bar{D}$ mesons ($\bar {D^0}$, $D^-$), the properties of the 
$D$ mesons ($D^0$,$D^+$) are more sensitive to the isospin asymmetry 
at high densities. On the other hand, the effects of strangeness 
fraction are found to be more dominant for the $\bar{D}$ mesons 
as compared to $D$ mesons and these modifications are observed 
to be particularly appreciable at high densities. We also study 
the mass modifications of the charmonium states $J/\psi$, $\psi(3686)$ 
and $\psi(3770)$ in the isospin asymmetric strange hadronic matter 
at finite temperatures and investigate the possibility of the
decay of the charmonium states into $D\bar D$ pairs in the
hot hadronic medium. The mass modifications of these charmonium states 
arise due to their interaction with the gluon condensates of QCD, 
simulated by a scalar dilaton field introduced to incorporate 
the broken scale invariance of QCD within the effective chiral model. 
The effects of finite quark masses are taken into account in the trace 
of the energy momentum tensor in QCD, while investigating the medium 
modification of the charmonium masses through the modification of 
the gluon condensate in the medium. We also compute the partial 
decay widths of the charmonium  states to the $D\bar {D}$ pairs 
in the hadronic medium. The strong dependence on density of the 
in-medium properties of the $D$, $\bar{D}$ and the charmonium states, 
as well as the partial decay widths of charmonium states to 
$D\bar {D}$ pairs, found in the present investigation, will be 
of direct relevance in observables like open charm enhancement 
as well as $J/\psi$ suppression in the compressed baryonic matter 
(CBM) experiments at the future Facility for Antiproton and Ion 
Research, GSI, where the baryonic matter at high densities 
is planned to be produced.
\end{abstract}
\maketitle

\def\bfm#1{\mbox{\boldmath $#1$}}

\section{Introduction}
The study of the medium modifications of hadrons is an important topic 
of research in strong interaction physics, which is of relevance in the 
heavy-ion collision experiments as well as in nuclear astrophysics.
The hadrons e.g. the vector mesons, the kaons and the $D$ mesons 
are made up of a light quark (u,d) and/or a light antiquark ($\bar u$,
$\bar d$). The mass of light quark is around 5-10 MeV whereas the the mass 
of the hadron e.g. proton is about 939 MeV. The large mass of the hadron 
is attributed to the spontaneous breaking of the chiral symmetry of QCD,
which leads to the population of the ground state of QCD with quark 
condensates. The masses of the hadrons are modified in the hadronic 
medium and these medium modifications can be linked with 
the medium modifications of the quark and/or gluon condensates of QCD.
The modifications of the hadron properties are of direct relevance
for the experimental observables of heavy ion collision experiments.  
The CERES experiment at CERN have observed an enhanced yield of 
dilepton pairs in heavy-ion reactions e.g. in S+Au collisions 
\cite{ceres}. The HELIOS \cite{helios} and DLS \cite{dls} collaborations  
also report high yield of dielectrons in the nucleus-nucleus collisions.
The above high observed yield of dileptons may be because of medium
modifications of the vector mesons and their decay to
dilepton pairs \cite{Brat1,CB99,vecmass,dilepton,liko}. The enhanced
yield in dileptons can also be a signature of formation of thermalized 
quark gluon plasma \cite{dk1}.
The dilepton pairs and photons observed in the heavy-ion collisions are
helpful in understanding the initial phase of collisions as
dileptons interact in the medium only electromagnetically
and most of the information of the collison zone remain undistorted 
\cite{dk2,bhatt1}. The properties of the kaons and antikaons 
in the hot and dense matter arising from heavy ion collision experiments
have been studied by KaoS collaboration \cite{K5,K6,K4,kaosnew}. 
The observed production and collective flow indicate that the kaons 
feel a repulsive interaction in the hadronic medium whereas antikaons 
undergo attractive interactions in the medium 
 \cite{CB99,cmko,lix,Li2001,cassing2003,kmeson,isoamss,isoamss2}.
The latter can lead to antikaon condensation in the neutron stars,
as was suggested by Kaplan and Nelson \cite{kaplan}.
 
The  study  of the mass modifications of  $D$ mesons is  
relevant in understanding their production as well as collective
flow in the heavy-ion collision experiments and have gotten 
considerable attention due to the experimental facilities, 
e.g., at Jefferson Lab, USA \cite{jlab} and at the upcoming
FAIR (Facility for Antiproton and Ion Research) project at GSI, 
Germany \cite{gsi}. In Jefferson Lab, CEBAF (Continuous Electron 
Beam Acclerator Facility) acclerator is used to get the continuous 
beam of electrons and these are scattered off from nuclei
to produce charm hadrons \cite{jlab}.
The CBM (Compressed Baryonic Matter) experiment at FAIR 
facility intends to explore the phase diagram of
 strongly interacting matter at high baryon densities
and moderate temperatures in heavy ion collisions
and the medium modification of the charm mesons are planned
to be investigated in these experiments. 
The PANDA experiment of FAIR project, GSI, in the experiment 
with the annihilation of antiprotons on the nuclei, also
intends to study $D$ meson and charmonium spectroscopy \cite{panda}.
On the theoretical side, the properties of $D$ mesons have been 
investigated in quark meson coupling (QMC) model  \cite{qmc}. 
In quark meson coupling model, $D(\bar D)$ mesons are assumed to be
bound states of a light quark (antiquark) and a charm antiquark (quark),
and they interact with the nucleons through the exchange of scalar 
and vector mesons. These studies suggest that 
the $D$ mesons undergo mass drop in the nuclear medium.
The QCD sum rule (QSR) approach has also been used to study the
medium modification of $D$ mesons in the nuclear medium 
\cite{arata,qcdsum08,weise}. In the QCD sum rule approach, the 
in-medium  properties of the hadrons, are related to the QCD ground 
state properties e.g. the quark and/or gluon condensates. 
The $D$ ($\bar{D}$) mesons, which are made up of one light 
antiquark (light quark), according to the QSR approach, 
undergo appreciable medium modification in their masses,
because of their interaction with the light quark condensates
in the nuclear medium.
The mass drop of $D$ mesons observed in the QCD sum rule approach 
\cite{arata} turns out to be similar as in QMC model \cite{qmc}.
The coupled channel approach has also been vastly used in the literature to
study the mass modifications of $D$ and $\bar{D}$ mesons in the nuclear 
medium at zero and finite temperatures \cite{ltolos,ljhs,mizutani6,mizutani8}.
The coupled channel approach has successfully been applied to understand the 
$\bar{K}N$ interactions in isospin I = 0 channel and dynamics of 
resonance $\Lambda(1405)$ \cite{kbarn}. The interest in the study of  
charm mesons have been triggered by the discovery of the charmed 
baryonic resonances by CLEO, Belle and BABAR collaborations
\cite{artuso,mizuk1,chistov,aubert1,aubert2,mizuk2}. However, 
it is a matter of interest to explore whether these resonances 
have $qqq$ structure or can be generated dynamically via meson baryon 
scattering processes \cite{ltolosdstar}. The study of the properties 
of $D$ mesons for the first time was studied within the coupled channel 
approach in Ref. \cite{ltolos} by assuming a separable baryon meson potential
and using the SU(3) symmetry in u,d and c quarks.
In Ref. \cite{ltolos} the resonance $\Lambda_{c}(2593)$ was produced 
in isospin I = 0 channel and the effects of the Pauli blocking of the 
nucleons, pion modification, baryon modification and
self dressing of $D$ mesons were taken into account
to investigate the medium modifications of the $D$ mesons. 
It was observed that there is only a small drop in the mass of $D$
mesons in the fully self-consistent calculations for the study
of the $D$ mesons in the nuclear medium \cite{ltolos}. In Ref. \cite{HL},
a coupled channel method was developed, which was based on SU(4) 
symmetry in $u, d, s$ and $c$ quarks. In this approach the free space 
scattering amplitudes were constructed through t-channel exchange 
of vector mesons and KSFR relations \cite{KSFR}. This method was later 
used to study the mass modifications of $D$ and $\bar D$ mesons in cold 
nuclear medium \cite{MK}. In Ref. \cite{mizutani6,mizutani8}, the original 
model developed by Hoffmann and Lutz was modified in several aspects
e.g. in terms of using cut-off regularization instead of dimensional 
regularization and was used to study the properties of $D$ mesons 
at zero \cite{mizutani6} as well as at finite temperatures 
\cite{mizutani8}. These studies generated a narrow resonance, 
$\Lambda_{c}(2593)$ in isospin I = 0 channel and a wide resonance, 
$\Sigma_{c}(2770)$ in I = 1 channel of $DN$ interactions. 
At zero temperature these resonances move to higher 
values of c.m. energies due to Pauli blocking effect, whereas 
the finite temperature effect leads to the movement of these resonances 
towards their free space position. The properties of $D$ and $D^{\ast}$ 
mesons have also 
been investigated using SU(8) spin flavor symmetry and lead to generation 
of rich spectrum of resonances, e.g., $\Lambda_{c}(2660), \Lambda_{c}(2941), 
\Sigma_{c}(2554), \Sigma_{c}(2902)$, with $J = 3/2$ and $\Sigma_{c}(2770)$ 
and $\Sigma_{c}(2823)$ with $J = 1/2$ \cite{ltolosdstar,garcia2,gamer,garcia1}.
The mass modification of the $D$ meson in the nuclear 
medium may also affect the properties of scalar charm and 
hidden charm resonances.  In Ref. \cite{tolosra}, it is observed that the 
spectral widths of resonances $D_{s0}(2317)$ and $X(3700)$ are modified 
considerably in the nuclear medium whereas the resonance $D(2400)$ 
is not affected much in the medium. 

The present paper is devoted to the study of medium 
modifications of $D$ and $\bar{D}$ mesons in isospin
asymmetric strange hadronic medium at finite temperatures.
In the present work, we also investigate the properties of the charmonium
states, $J/\psi, \psi(3686)$ and $\psi(3770)$ 
in isospin asymmetric strange hadronic matter at finite temperatures.
The investigation of the mass modifications of $D$ mesons and
charmonium states might be useful to understand anomalous
$J/\psi$ suppression observed by NA50 collaboration 
in Pb-Pb collisions \cite{NA501,NA50e,NA502}. The excited states 
of charmonium, considered as a major source of $J/\psi$
 mesons \cite{pAdata} could decay to $D\bar{D}$ pairs \cite{brat6} 
due to mass modifications of the $D$ and $\bar D$ mesons in the medium. 
The medium modifications of the $J/\psi$ mesons are observed to be small 
\cite{leeko} as compared to the $D$ meson mass modifications 
\cite{arata,haya1,friman}. According to QCD sum rules, this is attributed
to be due to the reason that the $D$ mesons interact in the nuclear medium 
through light quark condensates, which undergo appreciable modification 
in the medium. On the other hand, within the QCD sum rule approach,
the mass modifications of the $J/\psi$ mesons in the leading order
are due to the medium modifications of the gluon condensates, 
and the gluon condensates
have negligible modifications in the medium as compared to the medium 
modifications of the light quark condensates. The mass modification 
of the charmonium states in the medium 
have also been calculated from the self energy of $J/\psi$ due to
the $D$ meson loop \cite{leeko} as well as due to the $D^*$ meson loop 
\cite{krein1}. The mass modification of the charmonium states
due to the $D$-meson loop are observed to be smaller as compared 
to the effect due to the interaction with the gluon condensates,
whereas, the mass modification of $J/\psi$ due to the $D^*$ meson 
loop is observed to be large \cite{krein1}. 
In references \cite{satz} and \cite{blaiz}, 
$J/\psi$ suppression can be due to the formation of QGP since 
the charmonium dissociation rate is larger in QGP phase 
as compared to in the hadronic medium.
The additional $J/\psi$ suppression observed in SPS and RHIC 
energies can be due to co mover scattering \cite{sibirt,capella,vog}.
The $J/\psi$ mesons can also undergo absorption in the hadronic medium
\cite{zhang,brat5,elena}.
In the hadronic phase the $J/\psi$ mesons interact with the pions 
and $\rho$ mesons and form the charmed hadrons 
through reactions e.g. $\pi + J/\psi \rightarrow D^{\ast} + \bar{D}$ 
and therefore it is necessary to consider the medium modifications 
of mesons to explain the $J/\psi$ suppression phenomenon \cite{sibirt}.
In Ref. \cite{cassing} the phenomenon of $J/\psi$ suppression was studied
at SPS energies in covariant transport approach, whereas in 
Ref. \cite{zhang}, the $J/\psi$ suppression was investigated 
at RHIC energies in $Au + Au$ and $S + S$ collisions.
 It was observed that for the heavy nuclei both the plasma screening 
and the gluon scattering
are important and hadron absorption has only a minor effect, 
but for the light nuclei collisions the
effect of the plasma screening is observed to be negligible \cite{zhang}. 
 
In the present investigation, the isospin asymmetric hyperonic matter 
is described by a chiral SU(3) model \cite{paper3}. The chiral SU(3)
model is then generalized to SU(4) to study the in-medium
properties of the $D$ and $\bar D$ mesons. However, since the
chiral symmetry is broken for the SU(4) case by the large charm mass,
we use the SU(4) symmetry only to derive the interactions of the
D and $\bar D$ mesons in the hadronic medium, but use the
observed values of the charmed hadron masses and empirical values of
the decay constants \cite{liukolin}. In the present investigation,
the medium modifications of the charmonium states are considered 
through the medium modification of a scalar dilaton field $\chi$ 
which is introduced in the effective chiral model 
to mimic the the trace anomaly of QCD and is related
to the scalar gluon condensate of QCD \cite{amarvind}
The chiral SU(3) model used in the present work has already been used 
to investigate the medium modifications of the vector mesons 
\cite{hartree,kristof1} and the optical potentials of the
kaons and antikaons \cite{kmeson,isoamss,isoamss2}.
The properties of the $D$ mesons in symmetric/asymmetric nuclear
matter at zero and finite temperatures have also been studied 
\cite{amdmeson,amarind,amarvind} using the interactions of the 
charmed mesons, obtained by a generalization of the chiral SU(3) 
to SU(4). In the present investigation, we study the effects of the
strangeness on the properties of the $D$ and $\bar D$ mesons.
The properties of charmonium states in isospin asymmetric 
hot nuclear medium arising due to the medium modifications 
of the dilaton field using the  effective chiral model 
have been investigated \cite{amarvind,charmmass2} and the
present work studies the effects of strangeness on the
in-medium masses of the charmonium states.

The outline of the paper is as follows. In section 2, we give a brief 
introduction to the effective chiral $SU(3)$ model and its extension to 
SU(4) to study the medium modification of $D$ and $\bar{D}$ mesons in 
hot isospin asymmetric strange hadronic matter. In section 3, 
we write the interaction Lagrangian densities for $D$ and $\bar{D}$ mesons
with baryons and scalar mesons and derive the dispersion relations 
for the $D$ and $\bar{D}$ mesons. These dispersion relations are then
solved to obtain their optical potentials in the strange hadronic matter
at finite temperatures. In section 4, we investigate the in-medium masses 
of the charmonium states $J/\psi$, $\psi(3686)$ and $\psi(3770)$ 
arising from the medium modification of the scalar dilaton field, 
which is related to the modification of the gluon condensate in the 
hadronic medium. We also discuss the modifications of the  
size of the charmonium states in the hadronic medium. 
In section 5, we present the expressions of the partial 
decay widths of the charmonium states to $D\bar D$ pairs, obtained 
within the $3P0$ model \cite{friman,yaouanc,barnesclose}, and discuss 
our choice of parameters. In section 6 we discuss the results obtained 
for the medium modifications of the $D$($\bar D)$ mesons and the 
charmonium states in isospin asymmetric strange hadronic matter, 
as well as the effects of these modifications on the partial decay
widths of the charmonium states to the $D\bar {D}$ pairs. One observes 
nodes in the decay widths at certain densities when one considers the
mass modifications of the ($D\bar {D}$) mesons in the hadronic medium, 
as has been already observed in the literature \cite{friman}. However, 
appreciable modifications of these decay widths are found when the mass
modifications of the charmonium states are also taken into account. 
Finally, in section 7, we summarize the findings of the present 
investigation and discuss possible outlook.

\section{The effective chiral model}

The chiral $SU(3)$ model \cite{paper3} used in the present
investigation for the  study of the light hadrons is based on nonlinear 
realization of chiral symmetry \cite{weinberg,coleman,bardeen} and 
broken scale invariance \cite{paper3,hartree,kristof1}. A 
scalar dilaton field is introduced in the effective hadronic model 
to mimic the broken scale invariance of QCD \cite{paper3,amarvind,sche1}. 

The effective hadronic chiral Lagrangian contains the following terms
\begin{equation}
{\cal L} = {\cal L}_{kin}+\sum_W {\cal L}_{BW} + 
{\cal L}_{vec} + {\cal L}_{0} + {\cal L}_{scalebreak}+ {\cal L}_{SB}
\label{genlag}
\end{equation}
In Eq.(\ref{genlag}), ${\cal L}_{kin}$ is the kinetic energy term, 
${\cal L}_{BW}$ is the baryon-meson interaction term of meson of
type W, in which the baryons-spin-0 meson 
interaction term generates the baryon masses. ${\cal L}_{vec}$  describes 
the dynamical mass generation of the vector mesons via couplings to the 
scalar mesons and contains additionally quartic self-interactions of the 
vector fields. ${\cal L}_{0}$ contains the meson-meson interaction terms 
inducing the spontaneous breaking of chiral symmetry and the term
${\cal L}_{scalebreak}$ corresponds to the scale breaking potential
\cite{paper3,amarvind,sche1} given as
\begin{equation}
{\cal L}_{scalebreak}= -\frac{1}{4} \chi^{4} 
{\rm ln} \frac{\chi^{4}}{\chi_{0}^{4}} + \frac{d}{3} \chi^{4} 
{\rm ln} \Bigg( \frac{\left( \sigma^{2} - \delta^{2}\right)\zeta }
{\sigma_{0}^{2} \zeta_{0}} \Big( \frac{\chi}{\chi_{0}}\Big) ^{3}\Bigg), 
\label{scalebreak}
\end{equation}
where $\chi$, $\sigma$, $\zeta$ and $\delta$ are the scalar dilaton field,
non-strange scalar field, strange scalar field and the scalar-isovector
field respectively.
${\cal L}_{SB}$ describes the explicit chiral symmetry breaking. 
To study the hadron properties at finite densities,
we use the mean field approximation,
where all the meson fields are treated as classical fields. 
In this approximation, only the scalar and the vector fields contribute 
to the baryon-meson interaction, ${\cal L}_{BW}$ since for all the other 
mesons, the expectation values are zero. The coupled equations of motion 
of the scalar fields ($\sigma$, $\zeta$ and $\delta$) and the dilaton 
field, $\chi$, obtained from the mean field Lagrangian density,
are solved to obtain their values in the isospin asymmetric strange 
hadronic medium at finite temperatures. These medium dependent
scalar fields are then used to obtain the optical potentials 
for the $D$ and $\bar D$ mesons, as well as the in-medium masses
of the charmonium states for given values of density, temperature,
isospin asymmetry parameter, $\eta= -\frac{\sum_{i}  I_{3i} 
\rho_{i}}{\rho_{B}}$, with $\rho_i$ is the number density of 
the baryon of $i$-th type and the strangeness fraction,
$f_s= \frac{\sum_{i} s_{i} \rho_{i}}{\rho_{B}}$, where $s_{i}$ 
is the number of strange quarks of baryon $i$. The same procedure 
was used to study the properties of the $D$, $\bar D$ mesons 
as well as charmonium states for the isospin asymmetric hot 
nuclear matter ($f_s$=0) and the possibility of decay of the 
charmonium states to $D\bar D$ pairs 
in the hot nuclear medium was discussed in Ref. \cite{amarvind}. 
In the present work, we study also the effect of strangeness on 
the in-medium properties of these charmed mesons. As has already been 
mentioned, the medium modification of the masses of the charmonium states 
are investigated from the modification of the gluon condensates
in the hadronic medium \cite{leeko}, which are related 
to the medium modification of the dilaton field in the effective 
chiral model used in the present investigation \cite{amarvind}. 
The comparison of the trace of the energy momentum tensor
in QCD to the trace of the energy momentum tensor corresponding
to the scale breaking term of the effective chiral model given by
equation (\ref{scalebreak}) leads to the relation of the scalar 
gluon condensate to the dilaton field as \cite{charmmass2,cohen},
\begin{equation}
\theta_{\mu}^{\mu} = \langle \frac{\beta_{QCD}}{2g} 
G_{\mu\nu}^{a} G^{\mu\nu a} \rangle + \sum_i m_i \bar {q_i} q_i 
\equiv  -(1 - d)\chi^{4},
\label{tensorquark}
\end{equation}
where the second term in the trace accounts for the finite 
quark masses, with $m_i$ as the current quark mass for the quark
of flavor, $i=u,d,s,c$. 
The one loop QCD $\beta$ function is given as
\begin{equation}
\beta_{\rm {QCD}} \left( g \right) = -\frac{11 N_{c} g^{3}}{48 \pi^{2}} 
\left( 1 - \frac{2 N_{f}}{11 N_{c}} \right),
\label{qcdbeta}
\end{equation}
where $N_c=3$ is the number of colors and $N_f$ is the number of
quark flavors. Using equations (\ref{qcdbeta}) and (\ref{tensorquark}),
we obtain the scalar gluon condensate as related to the dilaton field, 
for $N_f=4$, as,
\begin{equation}
\langle \frac{\alpha_{s}}{\pi} 
G_{\mu\nu}^{a} G^{\mu\nu a} \rangle = \frac{24}{25}\left[ (1 - d)\chi^{4} 
+ \sum_i m_i \bar {q_i} q_i\right].
\label{chiglu}
\end{equation} 
The second term, $\sum_i m_i \bar q_i q_i$
can be identified to be the negative of the explicit chiral 
symmetry breaking term ${\cal L}_{SB}$ of equation (\ref{genlag})
\cite{charmmass2} and is given as
\begin{eqnarray}
\sum_i m_i \bar {q_i} q_i &=& - {\cal L} _{SB}  =  
\Big[ m_{\pi}^{2} 
f_{\pi} \sigma 
+ \left( \sqrt{2} m_{k}^{2}f_{k} - \frac{1}{\sqrt{2}} 
m_{\pi}^{2} f_{\pi} \right) \zeta  
+ \left( \sqrt{2} m_{D}^{2}f_{D} - \frac{1}{\sqrt{2}} 
m_{\pi}^{2} f_{\pi} \right) \zeta_{c} \Big],
\label{lsbcharm}
\end{eqnarray}
where, $\zeta_{c}$ is the scalar charm quark condensate $\bar{c}c$.
We thus see from the equation (\ref{chiglu}) that the scalar 
gluon condensate $\left\langle \frac{\alpha_{s}}{\pi} G_{\mu\nu}^{a} 
G^{\mu\nu a}\right\rangle$ is proportional to the fourth power of the 
dilaton field, $\chi$, in the limiting situation of massless quarks
\cite{amarvind}.

\section{$D$ and $\bar D$ mesons in hot isospin asymmetric strange 
hadronic matter}
In this section we study the $D$ and $\bar{D}$ mesons properties in isospin
asymmetric strange hadronic matter. As mentioned earlier, the medium 
modifications of the $D$ and $\bar{D}$ mesons arise due to their interactions 
with the nucleons, hyperons, and the scalar mesons.

The interaction Lagrangian density of the $D(\bar D)$ meson is given as
\begin{eqnarray}
{\cal L}_{int} & = & -\frac {i}{8 f_D^2} 
\Big [3\Big (\bar p \gamma^\mu p
+\bar n \gamma ^\mu n \Big) 
\Big(\Big({D^0} (\partial_\mu \bar D^0)
- (\partial_\mu {{D^0}}) {\bar D}^0 \Big )
+\Big(D^+ (\partial_\mu D^-) - (\partial_\mu {D^+})  D^- \Big )\Big )
\nonumber \\
& +&
\Big (\bar p \gamma^\mu p -\bar n \gamma ^\mu n \Big) 
\Big( \Big({D^0} (\partial_\mu \bar D^0) - (\partial_\mu {{D^0}}) 
{\bar D}^0 \Big )
- \Big( D^+ (\partial_\mu D^-) - (\partial_\mu {D^+})  D^- \Big )\Big )
\nonumber\\
&+& 2\Big((\bar{\Lambda}^{0}\gamma^{\mu}\Lambda^{0})
 \Big( \Big({D^0} (\partial_\mu \bar D^0)
 -(\partial_\mu {{D^0}}) {\bar D}^0 \Big)
+ \Big(D^+ (\partial_\mu D^-) - (\partial_\mu D^+)  D^- \Big) \Big)\nonumber\\
 &+& 2 \Big(\Big(\bar{\Sigma}^{+}\gamma^{\mu}\Sigma^{+}
 + \bar{\Sigma}^{-}\gamma^{\mu}\Sigma^{-}\Big)
 \Big(\Big({D^0} (\partial_\mu \bar D^0)
-(\partial_\mu {{D^0}}) {\bar D}^0 \Big)
+ \Big(D^+ (\partial_\mu D^-) - (\partial_\mu D^+)  D^- \Big)\Big)\nonumber\\
&+& \Big(\bar{\Sigma}^{+}\gamma^{\mu}\Sigma^{+}
 - \bar{\Sigma}^{-}\gamma^{\mu}\Sigma^{-}\Big)
 \Big(\Big({D^0} (\partial_\mu \bar D^0)
 -(\partial_\mu {{D^0}}) {\bar D}^0 \Big)
- \Big(D^+ (\partial_\mu D^-) - (\partial_\mu D^+)  
D^- \Big)\Big)\Big)\nonumber\\
&+&2\Big(\bar{\Sigma}^{0}\gamma^{\mu}\Sigma^{0}\Big)
 \Big(\Big({D^0} (\partial_\mu \bar D^0)
 -(\partial_\mu {{D^0}}) {\bar D}^0 \Big)
+ \Big(D^+ (\partial_\mu D^-) - (\partial_\mu D^+)  D^- \Big) \Big)\nonumber\\
 &+& \Big(\bar{\Xi}^{0}\gamma^{\mu}\Xi^{0}
 + \bar{\Xi}^{-}\gamma^{\mu}\Xi^{-}\Big)
 \Big(\Big({D^0} (\partial_\mu \bar D^0)
 -(\partial_\mu {{D^0}}) {\bar D}^0 \Big)
+ \Big(D^+ (\partial_\mu D^-) - (\partial_\mu D^+)  D^- \Big)\Big)\nonumber\\
&+&\Big(\bar{\Xi}^{0}\gamma^{\mu}\Xi^{0}
 - \bar{\Xi}^{-}\gamma^{\mu}\Xi^{-}\Big)
 \Big(\Big({D^0} (\partial_\mu \bar D^0)
 -(\partial_\mu {{D^0}}) {\bar D}^0 \Big)
- \Big(D^+ (\partial_\mu D^-) - (\partial_\mu D^+)  
D^- \Big)\Big)\Big ]\nonumber\\
&+& \frac{m_D^2}{2f_D} \Big [ 
(\sigma +\sqrt 2 \zeta_c)\big (\bar D^0 { D^0}+(D^- D^+) \big )
+\delta \big (\bar D^0 { D^0})-(D^- D^+) \big )
\Big ] \nonumber \\
& - & \frac {1}{f_D}\Big [ 
(\sigma +\sqrt 2 \zeta_c )
\Big ((\partial _\mu {{\bar D}^0})(\partial ^\mu {D^0})
+(\partial _\mu {D^-})(\partial ^\mu {D^+}) \Big )
  + \delta
\Big ((\partial _\mu {{\bar D}^0})(\partial ^\mu {D^0})
-(\partial _\mu {D^-})(\partial ^\mu {D^+}) \Big )
\Big ]\nonumber \\
&+& \frac {d_1}{2 f_D^2}(\bar p p +\bar n n +\bar{\Lambda}^{0}\Lambda^{0}+\bar{\Sigma}^{+}\Sigma^{+}+\bar{\Sigma}^{0}\Sigma^{0}
+\bar{\Sigma}^{-}\Sigma^{-}+\bar{\Xi}^{0}\Xi^{0}+\bar{\Xi}^{-}\Xi^{-})
  \big ( (\partial _\mu {D^-})(\partial ^\mu {D^+})\nonumber \\
&+&(\partial _\mu {{\bar D}^0})(\partial ^\mu {D^0})
\big )
+ \frac {d_2}{2 f_D^2} \Big [
\Big(\bar p p+\frac{1}{6}\bar{\Lambda}^{0}\Lambda^{0}
+\bar{\Sigma}^{+}\Sigma^{+}+\frac{1}{2}\bar{\Sigma}^{0}\Sigma^{0}\Big)
(\partial_\mu {\bar D}^0)(\partial^\mu {D^0})\nonumber \\
&+&\Big(\bar n n+\frac{1}{6}\bar{\Lambda}^{0}\Lambda^{0}
+\bar{\Sigma}^{-}\Sigma^{-}+\frac{1}{2}\bar{\Sigma}^{0}\Sigma^{0}\Big)
 (\partial_\mu D^-)(\partial^\mu D^+)\Big ]
\label{lddbar}
\end{eqnarray}

In Eq. (\ref{lddbar}), the first term is the vectorial Weinberg Tomozawa 
interaction term, obtained from the baryon-pseudoscalar meson
interaction Lagrangian given for the SU(4) case as follows. 
\begin{eqnarray}
{\mathcal L}_{\rm{WT}} & = & -\frac{1}{2}\sum_{i,j,k,l}
\bar B_{ijk} \,\gamma^\mu\,\Big( {({\Gamma_{\mu}})_l}^k\,B^{ijl}
+ 2\, {({\Gamma_{\mu}})_l}^j\,B^{ilk}\Big),
\label{WTtensor}
\end{eqnarray}
In the above, the baryons belong to the 20-plet representation 
and mesons belong to the 16-plet representation. The baryons are 
represented by the tensor
$B^{ijk}$, which are antisymmetric in the first two indices \cite{HL}. 
The indices $i,j,k$ run from one to four, where one can read off the
quark content of a baryon state by the identifications $1 \leftrightarrow u,
2 \leftrightarrow d, 3 \leftrightarrow s, 4 \leftrightarrow c$.
The baryon states are given as \cite{HL}, 
\begin{eqnarray}
\begin{array}{lll}
B_{121}=p,\;\; B_{122}=n,\;\; B_{213}=\frac2{\sqrt6}\Lambda^{0}, \\ 
B_{132}=\frac1{\sqrt2} \Sigma^0 -\frac1{\sqrt6}\Lambda^{0},\;\;\; 
 B_{231}=\frac1{\sqrt2}\Sigma^0+\frac1{\sqrt6}\Lambda^{0}, \\
 B_{232}=\Sigma^-,\; B_{311}=\Sigma^+, \;
B_{233}=\Xi^-,\; B_{313}=\Xi^0 \\
    \end{array}
\end{eqnarray}
where we have written down only the baryons containing the three light quarks,
u, d and s quarks.
The second term in equation (\ref{lddbar}) is the scalar meson exchange 
term, which is obtained from the explicit symmetry breaking term
\begin{equation}
\label{esb-gl}
 {\cal L}_{SB}  =  
     -\frac{1}{2} \Tr A_p \left(uXu+u^{\dagger}Xu^{\dagger}\right) 
\end{equation}
where, $A_{p}$ given as,
\begin{eqnarray}
A_p&=&1/\sqrt{2} {\mathrm{diag}}(m_{\pi}^2 f_{\pi},
 m_\pi^2 f_\pi,
 2 m_K^2 f_K 
-m_{\pi}^2 f_\pi, 2 m_D^2 f_D
-m_{\pi}^2 f_\pi),
\end{eqnarray}
 $X$ is the scalar meson multiplet \cite{amarind}. 
In the above, u is given as,
\begin{equation}
u  = exp\left[ \frac{i}{\sqrt{2}\sigma_{0}}M\gamma_{5}\right],
\label{piexp}
\end{equation}
with $M = \frac{1}{\sqrt{2}}\pi_{a}\lambda_{a}$
as the pseudoscalar meson multiplet \cite{amarind}. 
The next three terms of 
equation (\ref{lddbar}) ($\sim (\partial_\mu {\bar D})(\partial ^\mu D)$)
are the range terms. 
The first range term (with coefficient 
$\big (-\frac{1}{f_D}\big)$) is obtained from the kinetic energy term 
of the pseudoscalar mesons which is defined as,
\begin{equation}
\label{pikin}
 {\mathcal L}_{{\mathrm{1st range term}}} =  Tr (u_{\mu} X u^{\mu}X +X u_{\mu} u^{\mu} X) . 
\end{equation}
where, $u_{\mu}$ is given as,
\begin{equation}
u_{\mu} =-\frac{i}{2} \left[u^{\dagger}(\partial_{\mu}u) 
-u (\partial_{\mu}u^\dagger) \right]. 
\end{equation}
The range terms  $d_{1}$ 
and $d_{2}$ of equation (\ref{lddbar}) are
obtained from the expressions
\begin{eqnarray}
{\mathcal L}_{d_{1}} = \frac{d_{1}}{4}\sum_{i,j,k,l=1}^4\,
\bar B_{ijk} B^{ijk}{(u_\mu)}_{l}^{m}{(u^\mu)}_{m}^{l}
\label{d1tensor}
\end{eqnarray}
and
\begin{eqnarray}
{\mathcal L}_{d_{2}} =\frac{d_{2}}{2}\sum_{i,j,k,l=1}^4\,
\bar B_{ijk} {{(u_\mu)}_{l}}^{m}{{(u^\mu)}_{m}}^{k}B^{ijl} 
\label{d2tensor}
\end{eqnarray}
respectively.
We have not taken into account vector meson-pseudoscalar interactions 
in the present investigation as  we have retained only the leading
and next to leading order contributions and in the nonlinear realization
of chiral symmetry, such a term arises as a higher order contribution
\cite{borasoy}.

The  dispersion relations for the $D$ and $\bar{D}$ mesons are obtained 
by the Fourier transformations of equations of motion. These are given as 
\begin{equation}
-\omega^{2}+\vec{k}^{2}+m_{D}^{2}-\Pi\left(\omega,\vert\vec{k}\vert 
\right) 
= 0
\label{dispersion}
\end{equation}
where, $m_D$ is the vacuum mass of the $D(\bar D)$ meson and
$\Pi\left(\omega,\vert\vec{k}\vert\right)$ is the self-energy 
of the $D\left( \bar{D} \right) $ mesons in the medium.
The self-energy $\Pi\left( \omega , \vert\vec{k}\vert, \right) $ 
for the $D$ meson doublet $ \left( D^{0} , D^{+}\right) $ arising 
from the interaction of Eq.(\ref{lddbar}) is given as
\begin{eqnarray}
\Pi (\omega, |\vec k|) &= & \frac {1}{4 f_D^2}\Big [3 (\rho_p +\rho_n)
\pm (\rho_p -\rho_n) 
+2\big(\left( \rho_{\Sigma^{+}}+ \rho_{\Sigma^{-}}\right) \pm
\left(\rho_{\Sigma^{+}}- \rho_{\Sigma^{-}}\right) \big)\nonumber\\
&+&2(\rho_{\Lambda^{0}}+\rho_{\Sigma^{0}})
+( \left( \rho_{\Xi^{0}}+ \rho_{\Xi^{-}}\right) 
\pm 
\left(\rho_{\Xi^{0}}- \rho_{\Xi^{-}}\right)) 
\Big ] \omega \nonumber \\
&+&\frac {m_D^2}{2 f_D} (\sigma ' +\sqrt 2 {\zeta_c} ' \pm \delta ')
+ \Big [- \frac {1}{f_D}
(\sigma ' +\sqrt 2 {\zeta_c} ' \pm \delta ')
+\frac {d_1}{2 f_D ^2} (\rho_p ^s +\rho_n ^s\nonumber\\
&+&\rho_{\Lambda^{0}}^s+\rho_{\Sigma^{+}}^s+\rho_{\Sigma^{0}}^s
+\rho_{\Sigma^{-}}^s
+\rho_{\Xi^{0}}^s+\rho_{\Xi^{-}}^s)
+\frac {d_2}{4 f_D ^2} \Big ((\rho _p^s +\rho_n^s)
\pm   ({\rho} _p^s -{\rho}_n^s)+\frac{1}{3}{\rho} _{\Lambda^0}^s\nonumber\\
&+&({\rho}_{\Sigma^{+}}^s +{\rho} _{\Sigma^{-}}^s)
\pm    ({\rho} _{\Sigma^{+}}^s -{\rho}_{\Sigma^{-}}^s)
+{\rho} _{\Sigma^{0}}^s \Big ) \Big ]
(\omega ^2 - {\vec k}^2).
\label{selfd}
\end{eqnarray}
where the $\pm$ signs refer to the $D^{0}$ and $D^{+}$ mesons, 
respectively, and $\sigma^{\prime}\left( = \sigma - \sigma_{0}\right) $, 
$\zeta_{c}^{\prime}\left( = \zeta_{c} - \zeta_{c0}\right)$, and 
$\delta^{\prime}\left(  = \delta -\delta_{0}\right) $ are the 
fluctuations of the scalar-isoscalar fields $\sigma$, $\zeta_{c}$ and
the scalar-isoscalar field $\delta$ from their 
vacuum expectation values in the strange hyperonic medium. 
A non-zero value of the scalar isovector field, $\delta$
means the medium has isospin asymmetry and
 the vacuum expectation value of scalar isovector field $\delta$ will be zero.
Also, the fluctuation, $\zeta_{c}^{\prime}$, in the heavy charm quark
 condensate ($\zeta_{c} = \bar{c}c$)
from the vacuum value has been observed to be negligible \cite{roeder} 
and its contribution to the in-medium masses of $D$ and $\bar{D}$ mesons
will be neglected in the present investigation.
In equation (\ref{selfd}), $\rho_{i}$ and $\rho_{i}^{s}$ are 
the number density and 
the scalar density of the baryon of type $i$ with $i = p, n, 
\Lambda,\Sigma^{\pm,0}, \Xi^{-,0}$, and are defined as \cite{amarvind},
\begin{equation}
\rho_{i} = \gamma_{i} \int \frac{d^{3}k}{(2\pi)^{3}} 
 \left( \frac{1}{e^{\left( E_{i}^{*}(k) - \mu_{i}^{*}\right) /T} + 1} 
-  \frac{1}{e^{\left( E_{i}^{*}(k) + \mu_{i}^{*}\right) /T} + 1}\right), 
\label{vecdens}
\end{equation}
and 
\begin{eqnarray}
\rho_{i}^{s}& =& \gamma_{i}\int\frac{d^{3}k}{(2\pi)^{3}} 
\frac{m_{i}^{*}}{E_{i}^{*}(k)} 
 \Bigg ( \frac {1}{e^{({E_i}^* (k) -{\mu_i}^*)/T}+1}
+\frac {1}{e^{({E_i}^* (k) +{\mu_i}^*)/T}+1} \Bigg )
\label{scaldens}
\end{eqnarray}
respectively,
where $\gamma_i$=2 is the spin degeneracy factor.
  
For the  $\bar{D}$ meson doublet 
$\left(\bar{D}^{0},D^{-}\right)$, the expression for self-energy is
given as, 
\begin{eqnarray}
\Pi (\omega, |\vec k|) &= & -\frac {1}{4 f_D^2}\Big [3 (\rho_p +\rho_n)
\pm (\rho_p -\rho_n)
 +2\big(\left( \rho_{\Sigma^{+}}+ \rho_{\Sigma^{-}}\right)\pm 
\left(\rho_{\Sigma^{+}}- \rho_{\Sigma^{-}}\right) \big)\nonumber\\
&+&2(\rho_{\Lambda^{0}}+\rho_{\Sigma^{0}})
+( \left( \rho_{\Xi^{0}}+ \rho_{\Xi^{-}}\right) 
\pm 
\left(\rho_{\Xi^{0}}- \rho_{\Xi^{-}}\right))\Big ] \omega\nonumber \\
&+&\frac {m_D^2}{2 f_D} (\sigma ' +\sqrt 2 {\zeta_c} ' \pm \delta ')
 + \Big [- \frac {1}{f_D}
(\sigma ' +\sqrt 2 {\zeta_c} ' \pm \delta ')
+\frac {d_1}{2 f_D ^2} ({\rho}_p^s +{\rho}_n^s\nonumber\\
&+&{\rho}_{\Lambda^{0}}^s+{\rho}_{\Sigma^{+}}^s
+{\rho}_{\Sigma^{0}}^s+{\rho}_{\Sigma^{-}}^s
+{\rho}_{\Xi^{0}}^s+{\rho}_{\Xi^{-}}^s)
+\frac {d_2}{4 f_D ^2} \Big (({\rho}_p^s +{\rho}_n^s)
\pm   ({\rho}_p^s -{\rho}_n^s)+\frac{1}{3}{\rho}_{\Lambda^{0}}^s
\nonumber\\
&+&({\rho} _{\Sigma^{+}}^s +{\rho} _{\Sigma^{-}}^s)
\pm ({\rho}_{\Sigma^{+}}^s -{\rho}_{\Sigma^{-}}^s)
+{\rho} _{\Sigma^{0}}^s \Big ) \Big ]
(\omega ^2 - {\vec k}^2),
\label{selfdbar}
\end{eqnarray}
where the $\pm$ signs refer to the $\bar{D}^{0}$ and $D^{-}$ mesons, 
respectively. 
After finding the in-medium energies, $\omega(k)$ of $D$ and 
$\bar{D}$ mesons, we find their optical potentials using the 
expression
\begin{equation}
U(k) = \omega(k) - \sqrt{k^{2} + m_{D}^{2}}
\label{optpotential}
\end{equation}
where $m_{D}$ is the vacuum mass for the $D(\bar{D})$ meson.

\section{Charmonium states in hot asymmetric strange hadronic matter}
In this section, we outline the procedure which we use to study the
medium modification of the masses as well as the sizes of the 
charmonium states $J/\psi$, 
$\psi(3686)$ and $\psi(3770)$ in hot isospin asymmetric strange hadronic matter.
In the literature the masses of the charmonium states have been calculated 
using QCD sum rules through the medium modifications of the lowest dimension
gluon condensate operators which consists of the scalar and twist-2 
gluon condensates \cite{klingl}. These operators can, in turn, be
written in terms of the color electric field, $\langle \frac{\alpha_s}{\pi} 
{\vec E}^2\rangle$ and color magnetic field, $\langle\frac{\alpha_s}{\pi} 
{\vec B}^2\rangle$. However, in the non-relativistic limit, as
the Wilson coefficients for the magnetic field part vanish,
the lowest dimension gluon condensates can be written in terms
of the electric field part only and the mass shift of the 
charmonium states can be calculated as a second order Stark effect
in QCD, as has been studied in Ref. \cite{leeko}.
The expression for the mass shift of the charmonium state,
derived in the large charm mass limit is given as \cite{pes1}
\begin{eqnarray}
\Delta m_{\psi}  &=& -\frac{1}{9} \int dk^{2} \vert 
\frac{\partial \psi (k)}{\partial k} \vert^{2} \frac{k}{k^{2} 
/ m_{c} + \epsilon} 
\times \bigg ( 
\left\langle  \frac{\alpha_{s}}{\pi} E^{2} \right\rangle-
\left\langle  \frac{\alpha_{s}}{\pi} E^{2} \right\rangle_{0}
\bigg ).
\label{mass1}
\end{eqnarray}
In the above, $m_c$ is the mass of the charm quark, taken as 1.95 GeV 
\cite{leeko}, $m_\psi$ is the vacuum mass of the charmonium state 
and $\epsilon = 2 m_{c} - m_{\psi}$ is the binding energy. 
$\psi (k)$ is the wave function of the charmonium state
in the momentum space, normalized as $\int\frac{d^{3}k}{(2\pi)^{3}} 
\vert \psi(k) \vert^{2} = 1 $ \cite{leetemp}.
In the present investigation, the wave functions for the charmonium states
are taken to be Gaussian and are given as \cite{friman}
\begin{equation}
\psi_{N, l}(r) = {\rm { N}} \times Y_{l}^{m} (\theta, \phi)
(\beta^{2} r^{2})^{\frac{1}2{} l} exp^{-\frac{1}{2} \beta^{2} r^{2}}
L_{N - 1}^{l + \frac{1}{2}} \left( \beta^{2} r^{2}\right)
\label{wavefn}
\end{equation}
where $N$ is the normalization constant, 
$\beta^{2} = M \omega / \hbar$ characterizes the strength of the
harmonic potential, $M = m_{c}/2$ is the reduced mass of
the charm quark and charm anti-quark system, and $L_{p}^{k} (z)$
is the associated Laguerre Polynomial. As in Ref. \cite{leeko},
the oscillator constant $\beta$ is determined from the mean squared
radii $\langle r^{2} \rangle$ as 0.47$^{2}$ fm$^2$, 0.96$^{2}$ fm$^2$
and 1 fm$^{2}$ for the charmonium states $J/\psi$, $\psi(3686)$ 
and $\psi(3770)$, respectively. This gives the value for the parameter
$\beta$ as 0.52 GeV, 0.39 GeV and 0.37 GeV for $J/\psi$,
$\psi(3686)$ and $\psi(3770)$, assuming that these
charmonium states are in the 1S, 2S and 1D states respectively.

In the non-relativistic limit the color electric field part 
can be written in terms of the scalar gluon condensate as 
\cite{amarvind},
\begin{equation}
\left\langle \frac{\alpha_{s}}{\pi} E^{2} \right\rangle
=-\frac {1}{2} 
\left\langle \frac{\alpha_{s}}{\pi} 
G_{\mu\nu}^{a} G^{\mu\nu a}\right\rangle, 
\label{e2glu}
\end{equation}
which, in turn, gives the formula for the mass shift of the charmonium
states as \cite{amarvind} 
\begin{eqnarray}
\Delta m_{\psi} &=& \frac{1}{18} \int dk^{2} \vert 
\frac{\partial \psi (k)}{\partial k} \vert^{2} \frac{k}{k^{2} 
/ m_{c} + \epsilon} 
\times  \bigg ( 
\left\langle \frac{\alpha_{s}}{\pi} 
G_{\mu\nu}^{a} G^{\mu\nu a}\right\rangle -
\left\langle \frac{\alpha_{s}}{\pi} 
G_{\mu\nu}^{a} G^{\mu\nu a}\right\rangle _{0}
\bigg ).
\label{mass1}
\end{eqnarray}
The difference of the value of scalar gluon condensate in the medium 
and in the vacuum, using equations (\ref{chiglu}) and (\ref{lsbcharm}),
can be written as,
\begin{eqnarray}
&&\bigg ( 
\left\langle \frac{\alpha_{s}}{\pi} 
G_{\mu\nu}^{a} G^{\mu\nu a}\right\rangle -
\left\langle \frac{\alpha_{s}}{\pi} 
G_{\mu\nu}^{a} G^{\mu\nu a}\right\rangle _{0}
\bigg )
 =  \frac{24}{25}\bigg[  (1 - d)\left( \chi^{4} - \chi_{0}^{4}\right)
+  m_{\pi}^{2} 
f_{\pi} {\sigma}'
\nonumber\\
&+& \bigg( \sqrt{2} m_{k}^{2}f_{k} - \frac{1}{\sqrt{2}} 
m_{\pi}^{2} f_{\pi} \bigg ) {\zeta}' 
+ \bigg( \sqrt{2} m_{D}^{2}f_{D} - \frac{1}{\sqrt{2}} 
m_{\pi}^{2} f_{\pi} \bigg) {\zeta_{c}}' \bigg], 
\label{gludiff}
\end{eqnarray}
where $\sigma '$, $\zeta '$ and ${\zeta_c}'$ are the fluctuations
of the fields $\sigma$, $\zeta$ and $\zeta_c$ from their vacuum
values.
Using equation (\ref{gludiff}) in equation (\ref{mass1}), 
we calculate the mass shifts for the charmonium states.
As has already been mentioned in the last section, we neglect 
the fluctuation of the charm scalar field, $\zeta_c$, 
for the study of the in-medium properties of the
$D$ and $\bar D$ mesons in the present investigation. 
We shall not take this into account also in the study 
of the in-medium masses of the charmonium states
in the present study. We might note that 
when we do not consider the finite quark masses then 
the mass shift for the charmonium states is observed 
to be proportional to the shift in fourth power of the value 
of the dilaton field \cite{amarvind}.

\section{Decay widths of the charmonium states to $D\bar D$ pairs}

In this section, we compute the partial decay widths of the charmonium 
states to $D\bar D$ pairs in the hot isospin asymmetric strange 
hadronic medium, by accounting for the internal structures of the parent 
and outgoing mesons using the 3P0 model 
\cite{friman,yaouanc,barnesclose}.The medium modifications of the masses 
of the $D (\bar D)$ and the charmonium states as calculated in the 
present investigation, modify the decay widths of the charmonium states
to $D\bar D$ pairs in the medium. The charmonium state at rest decays 
to the  $D$ and $\bar D$ mesons and in the 3P0 model, this decay width 
is given as \cite{barnesclose}

\begin{equation}
\Gamma_{\psi \rightarrow D \bar D}
=2\pi \frac {p_D E_D E_{\bar D}}{M_\psi} 
|M|^2,
\label{decaywidth}
\end{equation}
where, $M$ is the matrix element for the decay of the parent charmonium
state to the $D\bar D$ pairs, $p_D$ is the magnitude of 
the 3-momentum of the $D$ ($\bar D$) meson when the charmonium state 
$\psi$ decays at rest and is given by
\begin{equation}
p_D=\Bigg (\frac{{M_\psi}^2}{4}-\frac {{m_D}^2+{m_{\bar D}}^2}{2}
+\frac {({m_D}^2-{m_{\bar D}}^2)^2}{4 {M_\psi}^2}\Bigg)^{1/2},
\label{pd}
\end{equation}
and $E_D$ and $E_{\bar D}$ are the energies of the outgoing $D$ and $\bar D$
mesons given as $E_D=({p_D}^2+{m_D}^2)^{1/2}$ and 
$E_{\bar D}=({p_D}^2+{m_{\bar D}}^2)^{1/2}$, with $m_D$ and $m_{\bar D}$
as the masses of the $D$ and $\bar D$ mesons.
In the isospin symmetric medium, the (almost) degeneracy in the masses of 
the $D^+$ and $D^0$ mesons, as well as in the masses of the $D^-$ 
and $\bar {D^0}$ mesons, leads to the partial decay widths of the 
charmonium states to $D^+D^-$ pair and $D^0\bar {D^0}$ pairs 
as approximately equal. However, in the isospin asymmetric medium,
the mass splitting between the masses of the $D^+$ and $D^0$,
as well as between the masses of the $D^-$ and $\bar {D^0}$,
lead to the partial decay widths for the channels when the charmonium state
decays to $D^+D^-$ and $D^0\bar {D^0}$ to be different.
The decay widths for the charmonium states $J/\psi$, $\psi(3686)$
and $\psi(3770)$ decaying to $D\bar D$ ($D^+D^-$ and $D^0\bar {D^0}$),
are given as \cite{friman}
\begin{eqnarray}
&& \Gamma (J/\psi \rightarrow D\bar D)
= \pi^{1/2} \frac {E_D E_{\bar D}\gamma^2}{2 M_{J/\psi}}
\frac { 2^8 r^3 (1+r^2)^2}{3 (1+2r^2)^5}x^3
\times \exp\Big(-\frac {x^2}{2(1+2r^2)}\Big),
\label{dwjpsi}
\end{eqnarray}
\begin{eqnarray}
&& \Gamma (\psi (3686) \rightarrow D\bar D)
= \pi^{1/2} \frac {E_D E_{\bar D}\gamma^2}{2 M_{\psi (3686)}}
\times \frac {2^7 (3+2r^2)^2(1-3r^2)^2}{3^2 (1+2r^2)^7}x^3
\nonumber \\ &\times & 
\Bigg (1+\frac {2 r^2(1+r^2)}{(1+2r^2)(3+2r^2)(1-3r^2)}x^2\Bigg)^2
\times  \exp\Big(-\frac {x^2}{2(1+2r^2)}\Big),
\label{dwpsi3686}
\end{eqnarray}
\begin{eqnarray}
&&\Gamma (\psi (3770) \rightarrow D\bar D)
= \pi^{1/2} \frac {E_D E_{\bar D}\gamma^2}{2 M_{\psi (3770)}}
\frac { 2^{11} 5}{3^2} \Big (\frac{r}{1+2r^2} \Big )^7
\times  x^3
\Bigg (1-\frac {1+r^2}{5(1+2r^2)}x^2\Bigg)^2
\exp\Big(-\frac {x^2}{2(1+2r^2)}\Big),
\label{dwpsi3770}
\end{eqnarray}
In the above, $r=\frac{\beta}{\beta_D}$
is the ratio of the harmonic oscillator strengths of the decaying 
charmonium state and the produced $D(\bar D)$-mesons, 
$x=p_D/\beta_D$ and $\gamma$ is a measure of 
the strength of the 3P0 vertex \cite{friman,barnesclose} . 
Few comments regarding the above expressions for the
decay widths may be in order. We might note here that 
the momentum dependence of the matrix element, $M$
of equation (\ref{decaywidth}) for the decay of charmonium 
state to $D\bar D$ pair arises from an overlap integral 
of the wave function of the decaying charmonium 
with the same of the $D$ and $\bar D$ mesons within 
the 3P0 model \cite{friman,yaouanc1}.
This overlap integral in general is a polynomial 
multiplied by a gaussian in the magnitude of 3-momentum,
$p_D$ of the produced $D(\bar D$) in the center of mass frame.
Therefore, the nodes of the wave functions in the coordinate
space can lead to nodes in the momentum, $p_D$ for the
decay amplitude. In particular, the polynomial part
in equation (\ref{dwpsi3686}) arises from the radial
part of the corresponding $2^3S_1$ wave function of
$\psi(3686)$ charmonium state, while the same in
equation (\ref{dwpsi3770}) arises from the orbital 
angular momentum part of the wave function of $1^3D_1$ wave function
of $\psi(3770)$ state \cite{friman}. Since the magnitude of the 
$D(\bar D)$, $p_D$ depends only upon the masses of the charmonium 
and the $D(\bar D)$ mesons, as can be seen from
equation (\ref{pd}), the medium modifications of the masses
of these mesons can lead to vanishing of the decay widths
at finite densities and temperatures. Such suppression
of the decay widths arising from the internal structure
of the mesons has already been observed for strong decays
of charmonium in the vacuum \cite{yaouanc1} as well as in the
medium \cite{friman}. In the present study, we also observe
such a suppression of the decay widths for the decay of 
the charmonium states in the isospin asymmetric hot strange 
hadronic matter, as shall be discussed in the following section.  

As has already been stated in the previous section, 
the strength of the harmonic oscillator potential for the charmonium 
state, $\beta$ is taken to be  0.52 GeV, 0.39 GeV and 0.37 GeV 
for $J/\psi$, $\psi(3686)$ and $\psi(3770)$ as fitted from their a
root mean squared radii, $\langle r^2 \rangle$ as 0.47$^2$ fm$^2$, 
0.96$^2$ fm$^2$ and 1 fm$^2$, respectively \cite{leeko}. 
The values of $\beta_D$ and $\gamma$ are taken to be 0.31 GeV 
and 0.35 respectively \cite{leeko}, so as to reproduce the decay 
width of $\psi(3770)$ to $D\bar D$ and partial decay widths of 
$\psi(4040)$ to $D\bar D$, $D\bar {D^*}$, $\bar D {D^*}$ 
and ${D^*}{\bar {D^*}}$ in vacuum \cite{friman,yaouanc1}.
With the values chosen for ratio of the strengths of the charmonium 
states and $D(\bar D)$ mesons and the 3P0 vertex strength, $\gamma$, 
the partial decay width of $\psi(3770)$
to $D\bar D$ turns out to be 28.65 MeV, with the partial decay widths
for the subchannels of $\psi(3770)\rightarrow D^+D^-$ and
$\psi(3770)\rightarrow D^0\bar {D^0}$ as 12.16 MeV and 16.49 MeV
respectively \cite{nakamura}.
With the strength of the outgoing $D$ and $\bar D$ mesons
taken as $\beta_D$= 0.31 GeV, the value of $r=\frac{\beta}{\beta_D}$ 
turns out to be 1.68, 1.26 and 1.19 for $J/\psi$, $\psi(3686)$
and $\psi(3770)$ respectively \cite{leeko} in the present
investigation. It may be noted here that in Ref. \cite{friman}, 
the decay widths of the charmonium states into $D\bar D$ pairs 
in isospin symmetric hadronic matter were computed within 
the 3P0 model, using the ratio of the strengths of the charmonium
to $D(\bar D)$ to be  $r=1.04$ for all the charmonium states.
The expressions for the partial decay widths 
for the charmonium states to the $D\bar D$ pairs as given by the 
equations (\ref{dwjpsi}), (\ref{dwpsi3686}) and (\ref{dwpsi3770})
account for the fact that the $D$ and $\bar D$ mesons can have 
different masses in the hadronic medium and these masses have been
calculated in the strange hadronic medium in the present investigation. 
We might note here that the earlier calculation of the decay widths 
of the charmonium states to $D\bar D$ pairs \cite{friman} had assumed 
the masses of the $D$ and $\bar D$ to be the same in the medium.
The partial decays widths are computed in the isospin asymmetric 
strange hadronic medium, accounting for the medium modifications 
of the charmonium states and the $D$ and $\bar D$ mesons, 
as calculated in the present investigation. The results for the 
medium modifications of these states as well as their effects 
on the partial decay widths of the charmonium states to $D\bar D$ 
pairs are discussed in the next section.

\section{Results and Discussions}

In this section, we present the numerical results of our investigation 
of the in-medium properties of $D$ and $\bar{D}$ mesons and of the
charmonium states in isospin asymmetric strange hadronic matter at 
finite temperatures, as well as the effects of these modifications
on the decay of the charmonium states to the $D \bar D$ pairs in the 
hadronic medium. The decay widths of the charmonium states
to the $D\bar D$ pairs have also been studied using 
the mass modifications of the charmonium states and
$D(\bar D)$ mesons calculated within the present model
as well as accounting for the 
internal structure of these mesons using the 3P0 model. 
Within the parameter set used in the chiral
effective model \cite{amarvind}, the vacuum values of the
non strange scalar isoscalar fields, $\sigma$ and strange
scalar isoscalar field, $\zeta$ and the dilaton field $\chi$ 
are $-93.3$ MeV, $-106.6$ MeV and $409.8$ MeV respectively.
The values, $g_{\sigma N} = 10.6$ and $g_{\zeta N} = -0.47$ 
are determined by fitting vacuum baryon masses. The other parameters 
fitted to the asymmetric 
nuclear matter saturation properties in the mean-field approximation 
are: $g_{\omega N}$ = 13.3, $g_{\rho p}$ = 5.5, $g_{4}$ = 79.7, 
$g_{\delta p}$ = 2.5, $m_{\zeta}$ = 1024.5 MeV, $ m_{\sigma}$ = 466.5 MeV 
and $m_{\delta}$ = 899.5 MeV. The values of the couplings of hyperons 
with the scalar fields are: $g_{\sigma\Lambda} = 7.52$, $g_{\zeta\Lambda} 
= 5.8$, $g_{\delta\Lambda} = 0$, $g_{\sigma\Sigma} = 6.13$, 
$g_{\zeta\Sigma} = 5.8$, $g_{\delta\Sigma^{+}} = 6.79$, 
$g_{\delta\Sigma^{-}} = -6.79$, $g_{\delta\Sigma^{0}} = 0$, 
$g_{\sigma\Xi} = 3.78$, $g_{\zeta\Xi} = 9.14$, $g_{\delta\Xi^{0}} = 2.36$
and $g_{\delta\Xi^{-}} = - 2.36$. The values of couplings of hyperons 
with the vector mesons are: $g_{\omega\Lambda}$ = $g_{\omega\Sigma}$ 
 = $\frac{2}{3}g_{\omega N}$, $g_{\rho\Sigma^{+}}$  
= $\frac{2}{3}g_{\omega N}$, $g_{\rho\Sigma^{-}}$  
= -$\frac{2}{3}g_{\omega N}$, $g_{\rho\Sigma^{0}}$  = 0,
$g_{\omega\Xi}$ =  $\frac{1}{3}g_{\omega N}$, $g_{\rho\Lambda} = 0$, 
$g_{\rho\Xi^{0}}$ = $\frac{1}{3}g_{\omega N}$,
$g_{\rho\Xi^{-}}$ = -$\frac{1}{3}g_{\omega N}$ 
$g_{\phi\Lambda}$ = $g_{\phi\Sigma}$   = $-\frac{\sqrt{2}}{3}g_{\omega N}$, 
$ g_{\phi\Xi}$   = $-\frac{2\sqrt{2}}{3}g_{\omega N}$ \cite{paper3}.
The coefficients $d_{1}$ and $d_{2}$ of the range term interactions
of the $D$ and $\bar D$ mesons are calculated from the empirical values 
of the $KN$ scattering lengths for $I = 0$ and $I = 1$ channels, 
and are given as $2.56/m_{K}$ and $0.73/m_{K}$, respectively 
\cite{isoamss2,amarvind}. 

\subsection{$D(\bar D)$ mesons in the hadronic matter}

The $D$ and $\bar{D}$ mesons properties in hot isospin asymmetric strange 
hadronic matter are modified due to their interactions with nucleons, 
hyperons, the scalar $\sigma$ meson, and scalar isovector $\delta$ meson.
As mentioned earlier, the non-strange scalar meson $\sigma$, the strange
scalar meson $\zeta$, the scalar isovector meson $\delta$ and the dilaton
field, $\chi$ are modified in the hadronic medium and their
values are obtained by solving the coupled equations of motion
of these scalar fields for given values of density, temperature,
isospin asymmetry parameter and strangeness fraction of the
hot hyperonic matter. The behavior of these scalar fields in 
isospin asymmetric nuclear medium ($f_s$=0) at zero as well as finite 
temperatures have been studied in detail in Ref. \cite{amarvind,amarind}.
In the present work, we shall focus on the effect of the strangeness 
fraction of the medium on these scalar fields. For T=0, density $\rho_B$
and a given isospin asymmetry $\eta$, the magnitude of the scalar field
$\sigma$ is observed to increase with increase in the
strangeness fraction in the medium at small densities, whereas 
at higher densities ($\rho_B > 3.5 \rho_0$), the magnitude of 
$\sigma$ is seen  drop with increase in $f_s$. On the other hand, 
the magnitude of $\zeta$ is observed to decrease with the strangeness 
in the medium. It is found that in isospin asymmetric
strange medium the magnitude of the scalar isovector field $\delta$
increases as we move from non strange medium ($f_{s} = 0$) to strange 
medium (finite $f_{s}$). However, at $\eta = 0$, the value of $\delta$ 
is zero because of isospin symmetry of the hadronic medium as expected,
since $\delta$ attains nonzero values only for the isospin asymmetric
hadronic matter. At given values for temperature, density and $\eta$,
the value of the dilaton field is observed to decrease with
the increase in the strangeness fraction of the medium. 

The effects of individual terms of Lagrangian density, given by 
equation (\ref{lddbar}), on the energies of $D$ and $\bar{D}$ mesons 
in isospin asymmetric hot nuclear matter have been studied in 
Refs. \cite{amarvind,amarind}. Here, we shall discuss the effect of 
strangeness fraction, $f_{s}$ of the medium 
on the energies of $D$ and $\bar{D}$ mesons arising due to the various
terms of the Lagrangian density given by equation (\ref{lddbar}). 
For a given value of density, $\rho_{B}$ and isospin asymmetry 
parameter, $\eta$, as we move from nuclear medium ($f_s$=0) 
to the hyperonic matter ($f_s \ne 0$), the attractive (repulsive) 
contribution to the in-medium energies of the $D({\bar D})$ mesons 
from the Weinberg-Tomozawa term is observed to become smaller.
The scalar meson exchange term is attractive for both the $D$ and 
$\bar D$ mesons. The increase in the magnitude of the $\sigma$ field
with strangeness at low densities, corresponds to a smaller value 
for $\sigma'$, the fluctuation of $\sigma$, ($\sigma '=sigma-\sigma_0$),
leading to a smaller drop in the masses of the $D$ and $\bar D$ mesons
as compared to their masses in nuclear matter ($f_s$=0). 
On the other hand, at high densities, there is a further
drop of the $D$ and $\bar D$ meson masses when the strangeness
fraction of the hadronic matter is increased.
Due to the behavior of the $\sigma$ field with strangeness
as described above, the magnitude of the first range term drops 
at small densities, whereas this term leads to an increase in the masses
of the D and $\bar D$ mesons at high baryon densities,
as we move from the nuclear medium ($f_s$=0) to hyperonic matter
(nonzero $f_s$). Also, as we increase the strangeness fraction
of the medium, the $d_{1}$ term is observed to become more attractive 
whereas the $d_{2}$ term becomes less attractive.
The attractive nature of $d_{1}$ term is found to be 
dominating over the first range term and the $d_{2}$ term.
Due to this reason, the contribution of the total range term
on the D and $\bar D$ masses is found to be more attractive
with the increase in the strangeness fraction of the medium.

\begin{figure}
\includegraphics[width=16cm,height=16cm]{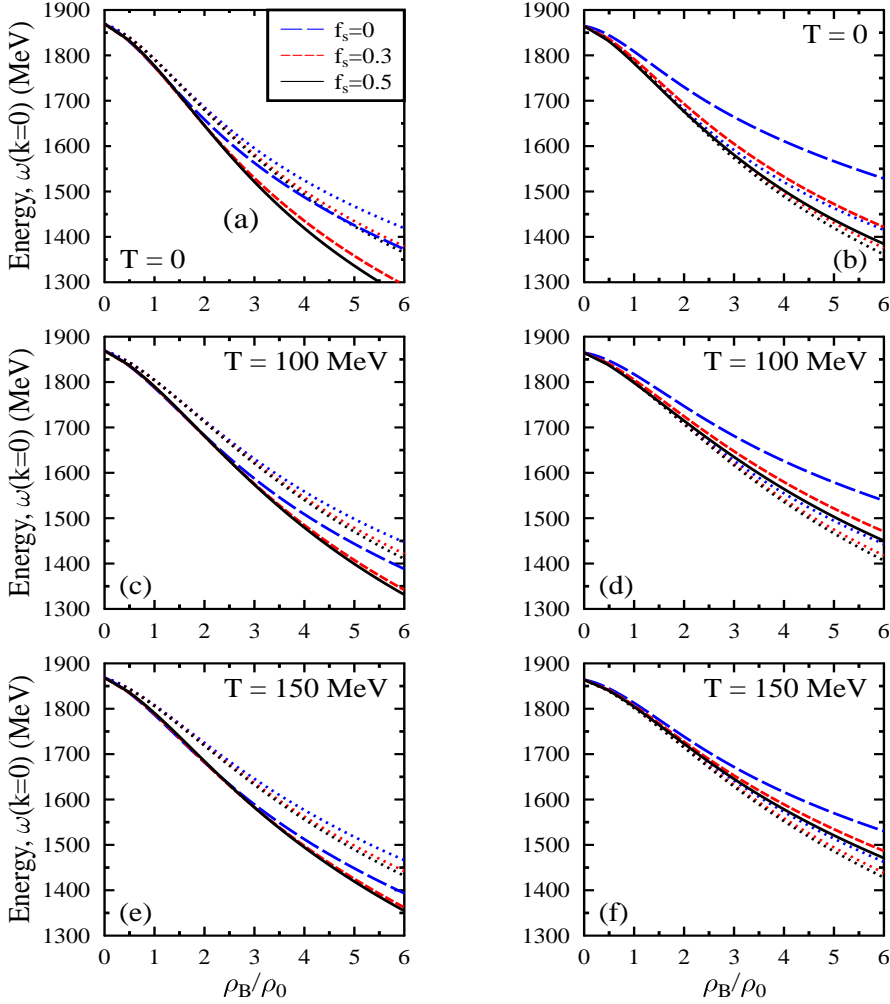}
\caption{(Color online) The energies of $D^{+}$ (subplots a, c, e) 
and $D^{0}$ (subplots b, d, f) mesons at zero momentum in isospin 
asymmetric ($\eta$=0.5) hyperonic medium for different temperatures, 
plotted as functions of baryon density in units of nuclear matter 
saturation density, $\rho_{B}/\rho_{0}$, for various values of the 
strangeness fraction, $f_s$. The results are compared with the case 
of isospin symmetric matter ($\eta = 0$) shown as dotted lines.}
\label{dmass}
\end{figure}

In Fig. \ref{dmass}, we show the variation of energies of 
$D^{+}$ [subplots (a), (c), (e)] and  $D^{0}$ [subplots (b), (d), (f)] 
mesons, at zero  momentum, with the baryon density, for temperatures 
T = 0, 100 and 150 MeV for different values of the strangeness fraction. 
In each subplot we show the results for 
isospin asymmetric medium with $\eta$=0.5 and compare these
energies with the values for symmetric hyperonic medium.
In isospin symmetric strange hadronic medium, for a given value of density, 
the energies of $D^{+}$ and $D^{0}$ mesons are observed to drop 
with increase in the strangeness fraction, $f_{s}$, of the medium
and this drop is seen to be larger at higher densities.
For example, at nuclear saturation density $\rho_{0}$ and for the 
isospin asymmetry parameter $\eta = 0$, as we move from $f_{s} 
= 0$ to $0.5$, the energies of both $D^{+}$ and $D^{0}$ mesons 
are observed to decrease by about $1$ MeV, whereas for $\rho_{B} 
= 4\rho_{0}$, there is seen to be a drop of about 31 MeV
from the value of 1524 MeV (1520 MeV) for $D^+ (D^0)$ meson,
as $f_{s}$ is changed from 0 to 0.5. In isospin asymmetric matter,
the drop in the D meson masses with $f_s$ is seen to be much larger
than for symmetric matter. For example, for $\eta = 0.5$, 
at $4\rho_{0}$ the energy of $D^{+}$ is observed to decrease 
by about $68$ MeV for $f_s$=0.5, from the value from 1487 MeV
at $f_s$=0. The energy of $D^{0}$ meson 
in isospin asymmetric strange hadronic medium ($\eta = 0.5$)
is observed to decrease by about 26 (109) MeV from the value 
of 1808 (1842) MeV at baryon density of $\rho_{0}$($4\rho_{0})$ 
as we move from $f_{s} = 0$ to $0.5$. 

For given values of isospin asymmetry and strangeness fraction, the energies 
of the $D$ mesons ($D^{0}, D^{+}$) are found to drop with increase in the 
density of the medium. In isospin symmetric nuclear medium ($\eta = 0$)
at zero temperature, the energy of $D^{+} (D^{0})$ meson at zero momentum 
is observed to decrease by about 77.3 (77.2) MeV and 345 (344) MeV 
from its vacuum value, at $\rho_{0}$ and $4\rho_{0}$ respectively. 
In isospin symmetric strange hadronic medium ($\eta = 0$), at the value of the
strangeness fraction $f_{s} = 0.5$, the energy of $D^{+} (D^{0})$ meson 
at $|\vec k|=0$ decreases by $78.3 (78.1)$ and $376.2 (375.5)$ MeV 
at $\rho_{0}$ and $4\rho_{0}$ respectively from its vacuum value.
The large drop in the masses of $D$ mesons in the hyperonic medium
is because of decrease in the mass of $D$ mesons with strangeness 
fraction of the medium, as has already been discussed. In isospin 
asymmetric strange hadronic medium ($\eta = 0.5$, $f_{s} = 0.5$), 
the energy of $D^{+} (D^{0})$ is observed to decrease drop 92 (82) MeV 
and 450 (362) MeV at densities of $\rho_{0}$ and $4\rho_{0}$ respectively 
from its vacuum value. 
    
The results stated above for $D$ mesons are for the strange hadronic medium
at zero temperature. As we move to the finite temperature of the strange 
hadronic medium the masses of the $D$ mesons increase and hence they have 
smaller drop from the vacuum values as compared to the zero temperature 
case. This has also been observed in nuclear medium calculation 
in Ref. \cite{amarvind}.
In isospin symmetric strange hadronic medium with strangeness 
fraction $f_{s} = 0.5$ and at a baryon density of $\rho_{0}(4\rho_{0})$ 
the mass of $D^{+}$ meson is observed to drop by
$72(357), 66(329)$ and $63(313)$ MeV from its vacuum value 
at temperatures T = 50, 100 and 150 MeV respectively.
In isospin asymmetric medium ($\eta = 0.5$) medium, at strangeness 
fraction, $f_{s} = 0.5$, and at a baryon density of
$\rho_{0}(4\rho_{0})$ the energy of $D^{+}$ meson is seen to 
decrease by $85(429)$,
 $79(390)$  and $78.6(374)$ MeV at temperatures
T = 50, 100 and 150 MeV respectively. For $f_s$=0.5 and $\eta$=0.5,
 at a density of $\rho_0(4\rho_0)$, the values of mass drop 
for $D^{0}$ mesons are observed to be $75(341)$, $66(300)$ and
 $60(286)$ MeV at temperatures T = 50, 100 and 150 MeV 
respectively. 
 
\begin{figure}
\includegraphics[width=16cm,height=16cm]{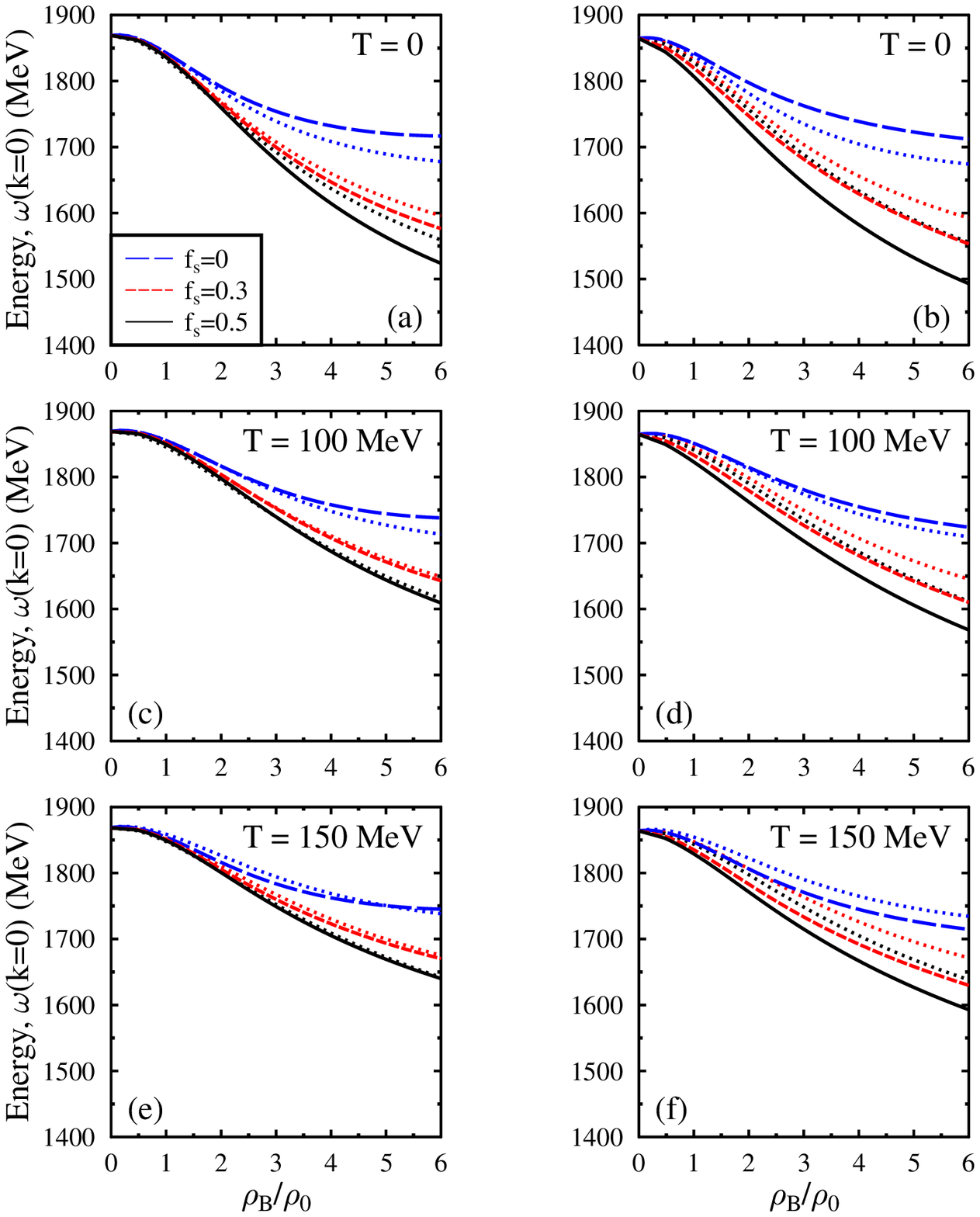}
\caption{(Color online)The energies of $D^{-}$ (subplots a, c, e) 
and $\bar{D^{0}}$ (subplot b, d, f) mesons at zero momentum in 
isospin asymmetric ($\eta$=0.5) hyperonic medium at different 
temperatures, plotted as functions of baryon density in units 
of nuclear matter saturation density, $\rho_{B}/\rho_{0}$, shown for 
various values of the strangeness fraction, $f_{s}$.
The results are compared with the case of isospin symmetric 
hadronic matter ($\eta = 0$) shown as dotted lines.} 
\label{dbarmass}
\end{figure}

In Fig. \ref{dbarmass}, we show the 
variation of the energies of $D^{-}$ [subplots (a), (c), (e)] 
and $\bar{D^{0}}$ [subplots (b), (d), (f)] mesons,
at zero  momentum, with density, for temperatures T = 0, 100 and  150 MeV.
In each subplot, we show the results for given values of the strangeness 
fraction, for the isospin asymmetric strange hadronic medium ($\eta$=0.5) 
and compared with the results for the symmetric matter ($\eta$=0).
At temperature T = 0, in isospin symmetric medium ($\eta = 0$), 
for a given value of density, as we move from 
$f_{s} = 0$ to $f_{s} = 0.5$, the energy of $D^{-}$ meson is observed to
decrease by 10 (71) MeV from the value of 1842 (1708) MeV at $f_s$=0,
at a density of $\rho_{0}$ ($4\rho_{0}$). The decrease in the energy 
of $\bar{D}$ mesons as a function of the strangeness fraction of the 
medium is because  of the drop in the mass with strangeness due to the 
Weinberg Tomozawa term as well as the total range term.
In the total range term, the attractive  $d_{1}$ term dominates over the
other range terms, as has already been discussed earlier. For the value 
of the isospin 
asymmetry parameter $\eta = 0.5$, the energy of $D^{-}$ meson decreases
by $7$ and $118$ MeV from the values of 1843 and 1732 MeV
at densities of $\rho_{0}$ and $4\rho_{0}$ respectively
when we increase the value of strangeness $f_0$ from 0 to 0.5. 
In isospin symmetric medium ($\eta = 0$), as we move from $f_{s} = 0$ 
to $0.5$, the energy of $\bar{D^{0}}$ meson is seen to decrease 
by about $10$ MeV and $71$ MeV from the values of 1838 and 1704.5 MeV
at $\rho_{0}$ and $4\rho_{0}$ respectively. At $\eta = 0.5$, 
the energy of $\bar{D^{0}}$ meson is seen to decrease by 35 (156) MeV 
from the value of 1842 (1739) MeV at a baryon density of 
$\rho_0$ ($4\rho_0$). As has already been mentioned, we thus observe 
in the present investigation, the masses of the $\bar D$ mesons 
are more sensitive to the strangeness of the medium as compared 
to the masses of D mesons, particularly at high densities.

Now we shall discuss the behavior of the $\bar{D}$ mesons 
in the strange hadronic medium at finite temperatures.
At finite temperatures, the drop in these masses are observed 
to be smaller than that of the zero temperature case.
In isospin symmetric medium($\eta = 0$), at baryon density
$\rho_{B} = \rho_{0}$ and strangeness fractions $f_{s} = 0(0.5)$, 
the mass of $D^{-}$ meson is seen to decrease by $20(30)$, 
$14(24)$ and $10(21)$ MeV from its vacuum value at values of 
temperature, T = 50, 100 and $150$ MeV respectively,
which may be compared to the zero temperature value of
27 (36.5) MeV for the mass drop.
For baryon density $\rho_{B} = 4\rho_{0}$, for $\eta$=0
and $f_s$=0(0.5), the values of the mass drop for $D^{-}$ mesons 
are modified to $147(210)$, $121(178)$ and $100(160)$ MeV at 
temperatures T = 50, 100 and $150$ MeV respectively,
which are smaller than the zero temperature values of mass drop
of 161.5 (228.5) MeV. In isospin asymmetric medium ($\eta = 0.5$), 
at baryon density $\rho_{B} = \rho_{0}$ and for the value of the 
strangeness fraction $f_{s} = 0(0.5)$, the mass of $D^{-}$ meson 
decreases by $20.5(26)$, $14(20)$ and $15(19.7)$ MeV at the value 
of the temperature T = 50, 100 and $150$ MeV respectively, whereas at
baryon density $\rho_{B} = 4\rho_{0}$ these values of mass
drop change to 128(229), $111(182)$ and 107(163) MeV respectively.
For $\bar{D^{0}}$ mesons, in isospin asymmetric medium ($\eta = 0.5$), 
at baryon density, $\rho_{0}$ and for the strangeness fraction 
$f_{s} = 0(0.5)$ the mass decreases by 18(51), $13(42)$ and 
$18(36)$ MeV at the values of the temperature,
 T = 50, 100 and $150$ MeV respectively from the vacuum value.
 For the same values for the isospin asymmetry and strangeness fraction,
the values of the mass drop changes to $119(259)$, $110(214)$ and 
$120(198)$ MeV, at baryon density, $\rho_B=4\rho_{0}$, 
at the value of the temperature, T = 50, 100 and 150 MeV respectively.
 
In the present investigation, we observe that the effect of strangeness 
fraction of the medium is to decrease the energies of the $D$ and 
$\bar{D}$ mesons. However, the $\bar{D}$ mesons 
are observed to undergo larger drop in their masses with increase
in the strangeness fraction of the medium as compared to the masses
of the D mesons. The difference in energies of $D$ and $\bar{D}$
mesons is due to the opposite sign in the contribution 
from the Weinberg Tomozawa term. With inclusion of hyperons in the medium, 
this term becomes less repulsive (attractive) for $\bar{D}$ (D)
mesons leading to an decrease (increase) in the energy of 
$\bar{D}$(D) mesons arising due to this term, as we increase $f_{s}$. 
This observed behavior of the contribution to the energy of
D($\bar D$) mesons arising from the Weinberg-Tomozawa term can be 
understood in the following manner. For example, for a given density,
for isospin symmetric nuclear matter, the expression in the bracket 
in the Weinberg-Tomozawa term in the self-energies of the D and 
$\bar D$ mesons given by equations (\ref{selfd}) and (\ref{selfdbar}) 
becomes equal to 3$\rho_B$. However, in the presence of hyperons,
but in isospin symmetric matter, the expression in the bracket
in this term becomes equal to $[3\rho_B-(\rho_\Lambda
+\rho_{\Sigma^+} +\rho_{\Sigma^-} +\rho_{\Sigma^0}+\rho_{\Xi^-}
+\rho_{\Xi^0})]$, which is less than 3$\rho_B$. Hence, for a given
baryon density, the magnitude of the Weinberg-Tomozawa term 
is observed to decrease with increase in the strangeness fraction
in the medium. The contributions to both the 
D and $\bar D$ mesons from the scalar exchange term as well as 
the range terms are the same. The scalar meson exchange term  
is seen to decrease the energies of $D$ and $\bar{D}$ mesons, with
increase in the strangeness in the medium, at densities
up to about 3.5 $\rho_0$ and leads to an increase at higher densities.
However, the modification in the energies of the D and $\bar D$ mesons
due to increase in the strangeness arising from the scalar exchange
term is seen to be very small. 
On the other hand, the total range term is observed to decrease the energies
of the D and $\bar D$ mesons appreciably with increase in the strangeness 
fraction of the medium, dominantly due to the attractive $d_1$ term. 
Hence, with increase in the strangeness fraction of the medium,
the Weinberg Tomozawa term as well as the total range term decrease 
the energy of $\bar{D}$ mesons, whereas for the D mesons, the Weinberg 
Tomozawa term leads to an increase in the energy with strangeness
and hence partly compensates the decrease due to increase in $f_s$
arising from the total range term. This leads to a larger drop
of the mass of the $\bar D$ mesons as compared to the D mesons
with increase in the strangeness in the medium. 

In the present calculations, we observe a drop
in the masses of the $D$ as well as $\bar D$ mesons.
The mass drop is seen to be larger in the presence of hyperons
in the medium, as can be observed from Figs. \ref{dmass} 
and \ref{dbarmass}. A drop in the masses of the $D$ mesons 
was also observed by the recent calculations in the coupled channel 
approach based on heavy quark symmetry \cite{garcia1}.
However, the drop in the $D$ meson mass in Ref. \cite{garcia1}
is seen to be  much smaller than the findings of the present 
investigation. The possibility of the formation of D-mesic nuclei 
arising from an attractive interaction of D-mesons in the
nuclear medium has also been investigated in Ref. \cite{garcia1}.
In the present investigation, for symmetric nuclear matter at zero
temperature, we obtain the mass shifts of about -77 and -27 MeV for the
$D^+$ and $D^-$ mesons. The mass shift of the $D$ meson may be
compared with the value of $-60$ MeV from calculations 
using quark meson coupling (QMC) model and of $-50$ MeV 
using the QCD sum rule approach \cite{arata}.
In Ref.\cite{MK}, the in-medium properties of open charm 
$D^{+}$ and $D^{-}$ mesons have been studied
in a self-consistent manner using coupled channel approach in
cold symmetric nuclear medium. The mass shifts of both the $D^+$ and
$D^-$ were observed to be repulsive \cite{MK}, with values of
30 and 18 MeV at nuclear matter saturation density.

\begin{figure}
\includegraphics[width=16cm,height=16cm]{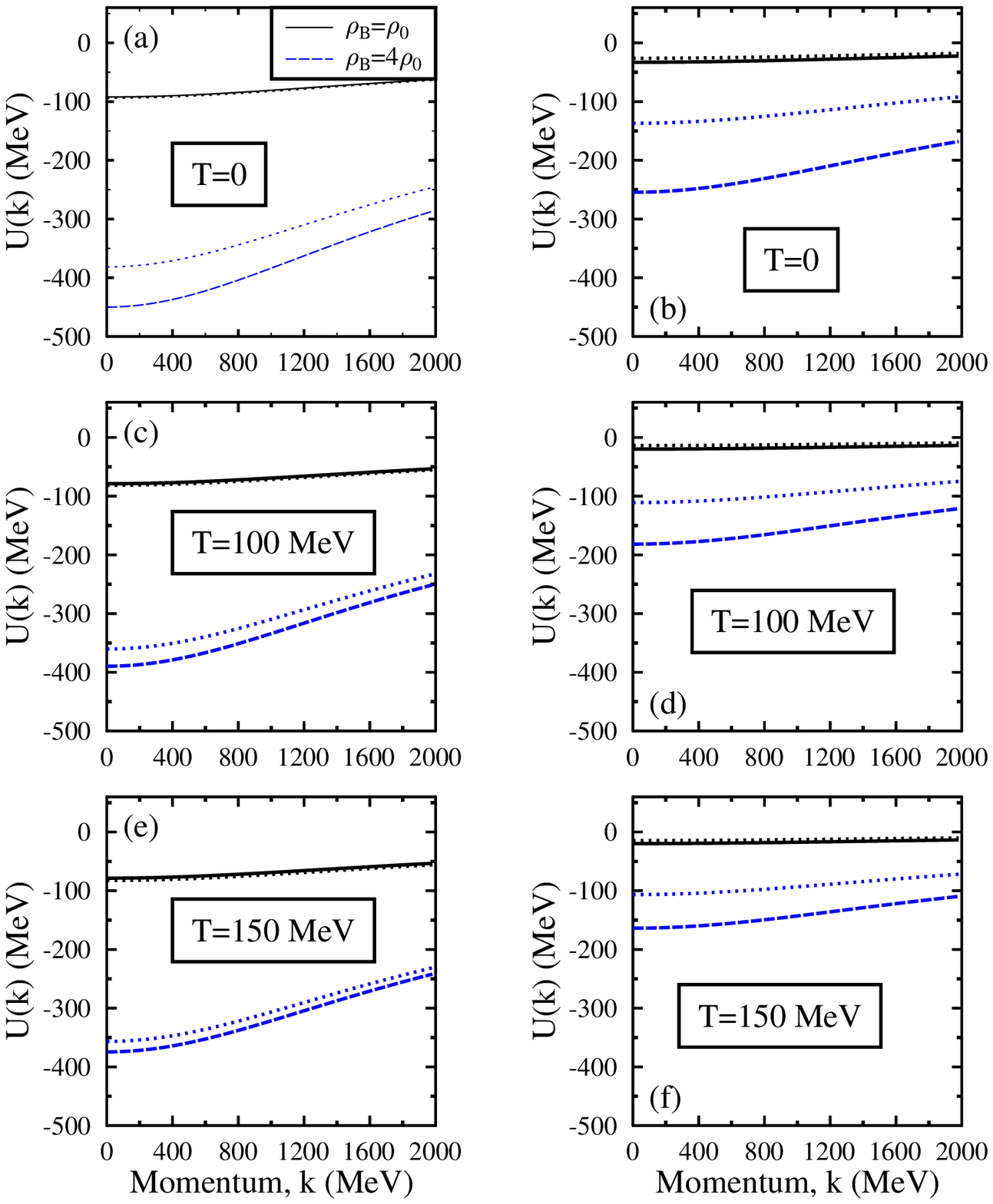}
\caption{(Color online) The optical potentials 
of the $D$ mesons (subplots (a), (c) and (e) are for $D^{+}$ 
mesons and (b), (d), and (f) are for $D^{0}$ mesons), plotted
as functions of momentum, k (MeV), at baryon densities of
$\rho_{B} = \rho_{0}$ and $4\rho_0$, for $\eta$=0.5 and $f_s$=0.5.
The results are compared with the case of $f_s = 0$ 
shown as dotted lines.} 
\label{optpotdk}
\end{figure}
In the above, we have investigated the energies of the D and $\bar D$
mesons at zero momentum in the hot asymmetric strange hadronic medium. 
We then study the effect of the finite momentum on the in-medium energies
of the $D$ and $\bar{D}$ mesons. In Fig. \ref{optpotdk},
we show the optical potentials of $D$ 
($D^{+}$ and $D^{0}$) mesons as functions of momentum, $\vert\vec{k}\vert$, 
at baryon densities, $\rho_{0}$ and $4\rho_{0}$ for $\eta$=0.5
and $f_s=0.5$ and compared to the situation for zero strangeness
in the medium. The medium modification of the masses of the D mesons 
are reflected in their optical potentials. The larger drop of the
mass of the $D^+$ meson as compared to the $D^0$ meson 
is reflected in their optical potentials and the strangeness is 
observed to give a larger value for the magnitude of the optical potentials
which reflects the fact that the mass drop of D mesons is larger 
with increase in strangeness in the medium as has already been 
illustrated in Fig. \ref{dmass}. 

\begin{figure}
\includegraphics[width=16cm,height=16cm]{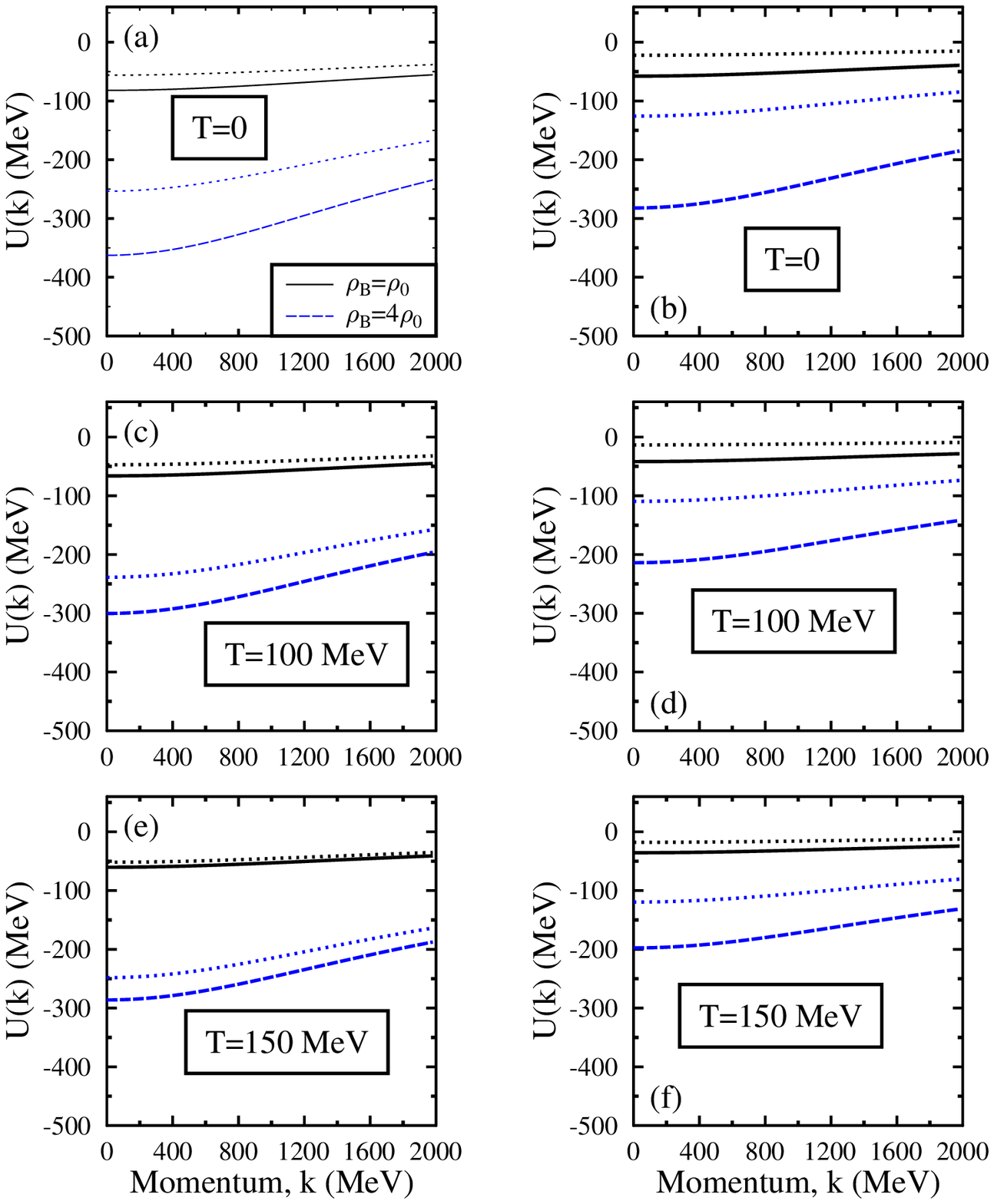}
\caption{(Color online)The optical potentials 
of the $\bar D$ mesons (subplots (a), (c) and (e) are for $D^{-}$ 
mesons and (b), (d), and (f) are for $\bar {D^0}$ mesons) plotted
as functions of momentum, k (MeV), at baryon densities of
$\rho_{B} = \rho_{0}$ and $4\rho_0$, for $\eta$=0.5 and $f_s$=0.5.
The results are compared with the case of $f_s = 0$ shown as 
dotted lines.} 
\label{optpotdbark}
\end{figure}
Fig. \ref{optpotdbark} shows the optical 
potentials of $\bar{D}$ mesons as functions of momentum, 
$\vert\vec{k}\vert$ at baryon densities $\rho_{0}$ and $4\rho_{0}$, 
which reflect the medium modifications of the
masses of the $\bar D$ mesons as shown in Fig. \ref{dmass}.
The optical potentials of $\bar{D}$ mesons also reflect 
the larger sensitivity of the mass of the $\bar {D^0}$ meson
to the strangeness content of the medium as compared to the
mass of the $D^-$ meson. The dependence of the optical potentials 
of $\bar {D^0}$ meson on strangeness of the medium is seen to be 
less at finite temperatures as compared to the case of zero 
temperature. This is related to the fact that the masses of 
the $D$ and $\bar D$ mesons have smaller drop in the medium 
as we increase the temperature. The momentum dependence of 
the optical potentials of the D and $\bar D$ mesons are observed 
to be small for small values of k. This is due to the reason that 
when $k^2$ is small as compared to ${m_D}^2$ or ${{m_D}^*}^2$, 
the optical potential given by equation (\ref{optpotential}) can be 
approximated as $({m_D}^*-{m_D})\big(1-k^2/(2{m_D}^* {m_D})\big)$, 
where the second term with the momentum dependence is small compared to 1
and hence for small k, the momentum dependence is observed to be small
for the optical potential. However, one observes the momentum dependence
to be larger for higher values of k, as can be seen from Figs.
\ref{optpotdk} and \ref{optpotdbark}.

Due to the different masses of the $D^+$ and $D^0$ mesons, as well as
of the $D^-$ and $\bar {D^0}$ mesons, in the isospin asymmetric hadronic 
medium, and due to the modifications of the masses of the charmonium 
states, the production cross sections, yield and the collective flow 
are expected to be different for the $D^+$ and $D^0$, as well as, 
for the $D^-$ and $\bar {D^0}$ mesons in the isospin asymmetric 
hadronic environment. 
The present investigation shows a stronger isospin dependence 
of the masses of the $D^+$ and $D^0$, as compared to the masses 
of the $D^-$ and $\bar {D^0}$ mesons, particularly at high densities 
and this is observed to be more dominant at higher values of the 
strangeness fraction. This should show up in the experimental observables
of the ratios of $D^+/D^0$ as well as $D^-/\bar {D^0}$ in the production 
cross-section as well as in the parameter, $v_2$ for the collective 
flow for the ratio $D^+/D^0$ from the high density matter resulting 
from compressed baryon matter (CBM) experiment at the FAIR project 
in the future facility at GSI. However,
the $D$ and $\bar D$ mesons, due to the presence of the light antiquark 
(quark) can undergo substantial decay within the hadronic environment,
and this could make it difficult to get clean signals from the
experimental data from the heavy ion collision experiments. 

\begin{figure}
\includegraphics[width=16cm,height=16cm]{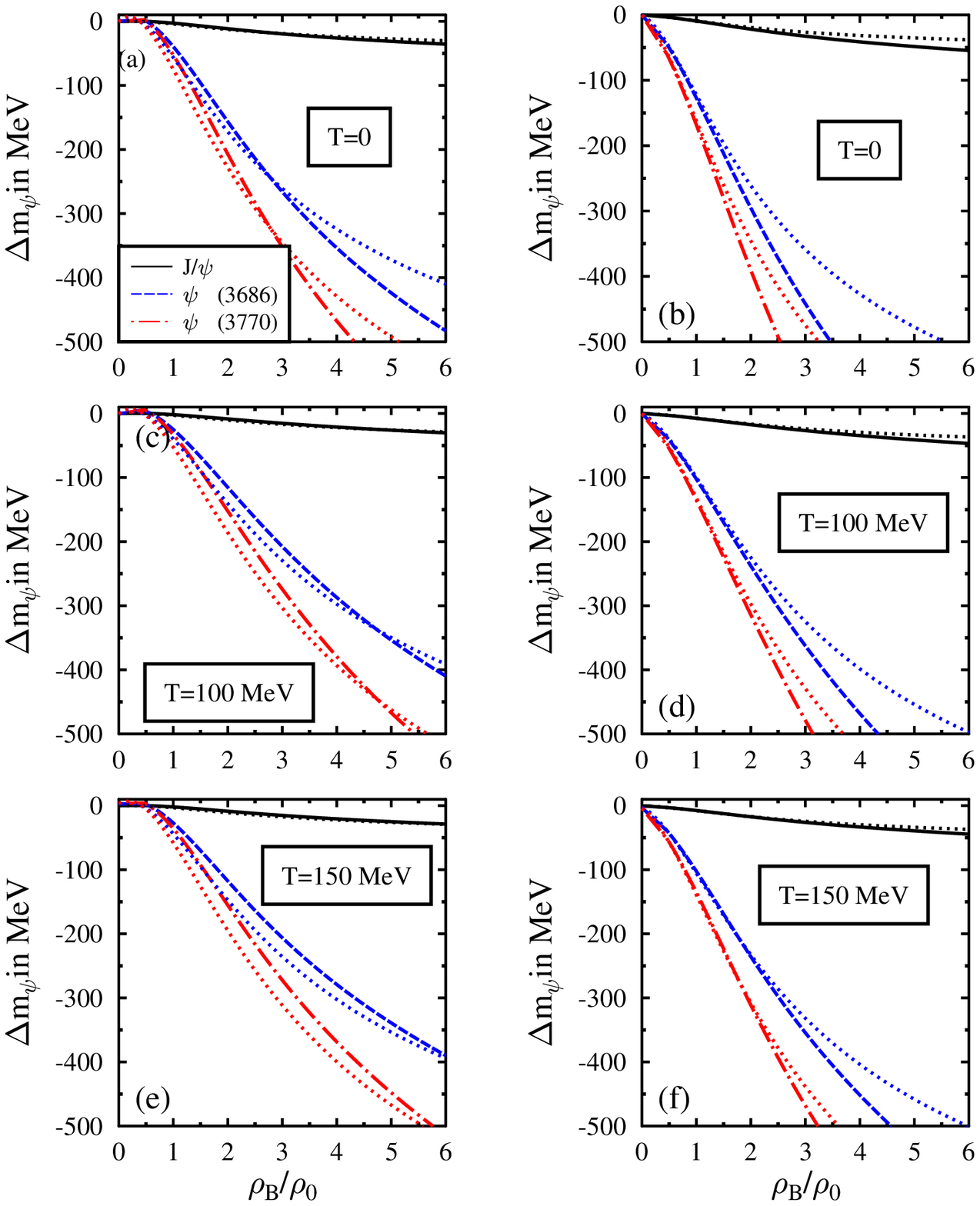}
\caption{(Color online) The mass shifts of the charmonium states,
$J/\psi$, $\psi (3686)$ and $\psi(3770)$ mesons in isospin
asymmetric hyperonic matter ($\eta$=0.5 and $f_s$=0.5), plotted
as functions of baryon density in units of nuclear matter saturation 
density, $\rho_{B}/\rho_{0}$, for values of temperature, T=0, 100
and 150 MeV. The mass shifts are shown in subplots (a), (c) and 
(e), when the gluon condensate in the medium is calculated 
by accounting for the finite quark mass term in the trace of the
energy momentum tensor, and in subplots (b), (d) and (f), 
when the quark mass term is neglected. The results are compared 
with the case of asymmetric nuclear matter ($f_s = 0$), 
shown as dotted lines.
} 
\label{charmdelm}
\end{figure}

\begin{figure}
\includegraphics[width=16cm,height=16cm]{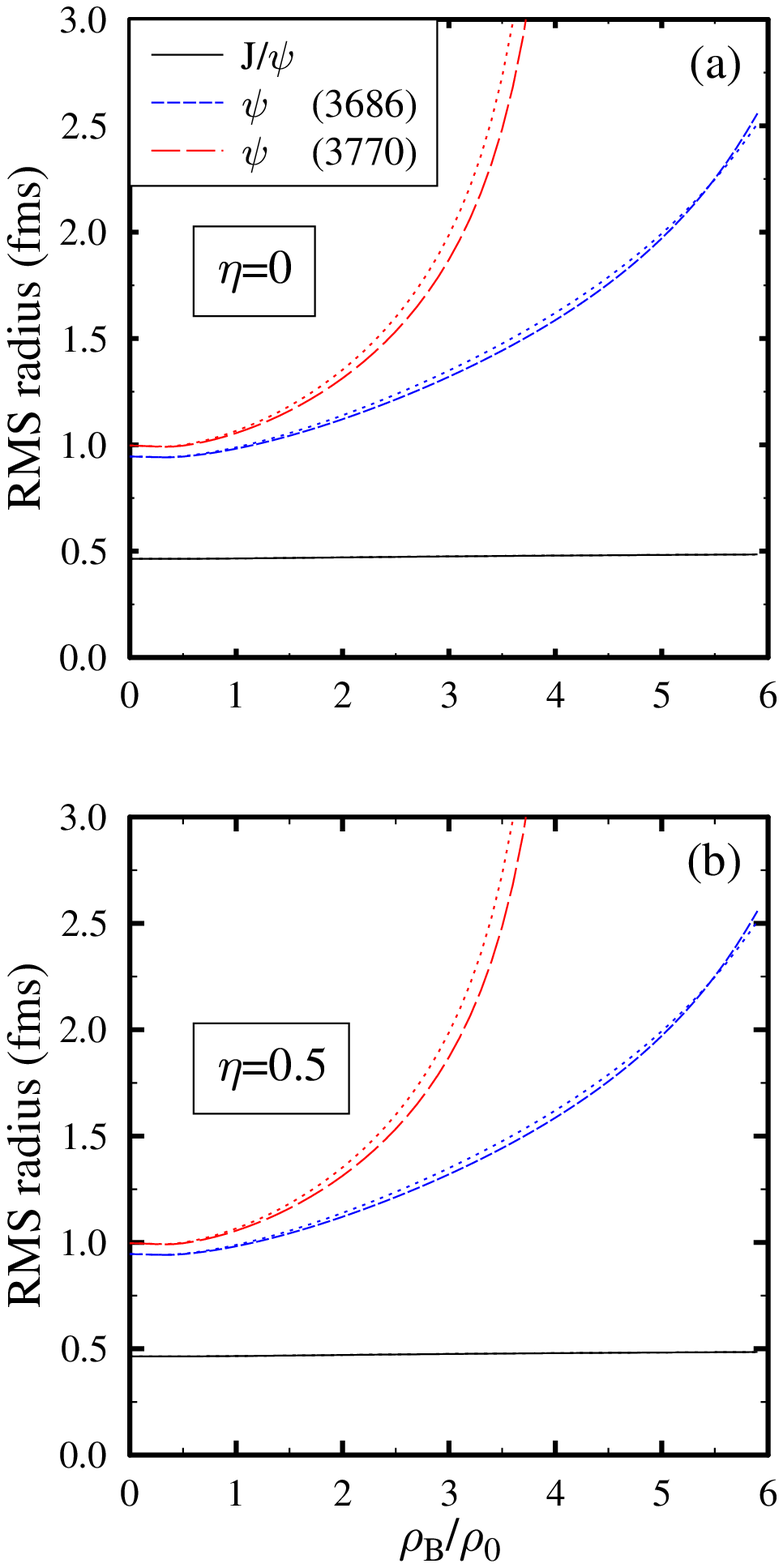}
\caption{(Color online) The rms radii of the charmonium states,
$J/\psi$, $\psi (3686)$ and $\psi(3770)$ plotted as functions 
of baryon density in units of nuclear matter saturation density, 
$\rho_{B}/\rho_{0}$, for $f_s$=0.5 and zero temperature. 
These have been shown in subplots (a) and (b) for the isospin
symmetric ($\eta$=0) and isospin asymmetric hyperonic matter
(with $\eta$=0.5). The results are compared with the case of 
asymmetric nuclear matter ($f_s = 0$), shown as dotted lines.
} 
\label{rmscharm}
\end{figure}

\subsection{Charmonium states in the hadronic matter}

We then investigate how the behavior of the dilaton field $\chi$ 
in the hot asymmetric strange hadronic matter affects the in-medium 
masses of the charmonium states $J/\psi$, $\psi(3686)$ and $\psi(3770)$. 
In Fig. \ref{charmdelm}, we show the shifts of the masses of the 
charmonium states $J/\psi$, $\psi(3686)$ and $\psi(3770)$ from their 
vacuum values, in the isospin asymmetric hyperonic matter 
($\eta$=0.5, $f_s$=0.5). These are plotted as functions of the 
baryon  density for
values of temperature, T = 0, 100 and 150 MeV. For these
temperatures, in the subplots (a), (c) and (e) the mass shifts 
in the charmonium states are shown when the gluon condensate 
in the medium is calculated with the finite quark mass term 
taken into account in the trace of the energy momentum tensor 
and the subplots (b), (d) and (f) show the results while neglecting 
the quark mass term.
The values of the mass shift of $J/\psi$ for the isospin symmetric medium
at zero temperature are found to be $-4.35 (-2.96)$ MeV and 
$-26.37 (-25.95)$ MeV for $f_s$=0(0.5) at $\rho_B=\rho_0$ and 
4$\rho_0$ respectively, 
and these values are modified to $-4.16 (-2.9)$ MeV and $-23.9 (-26.06)$ MeV 
for $\eta$=0.5. These mass shifts are for the situation when the
finite quark mass term is taken into account in the trace anomaly
and hence the gluon condensate which is calculated is a combined
effect of the modifications of the scalar dilaton field as well
as the quark condensates through the scalar $\sigma$ and $\zeta$ fields.  
One notices that the effects of temperature
are small on the masses of the charmonium states.
In isospin symmetric hadronic medium ($\eta = 0$) at zero temperature, 
at baryon density, $\rho_{B} = \rho_{0}$ and the value of the strangeness 
fraction $f_{s}=0(0.5)$, the mass shifts for the charmonium states, 
$\psi(3686)$ and $\psi(3770)$ are observed to be $-59(-40)$ MeV and
$-78.5(-54)$ MeV respectively, and these values are modified to
$-365.4 (-352)$ MeV and $-483.4 (-466)$ MeV for $\rho_B=4\rho_0$.
One observes that the effects of the strangeness, isospin asymmetry
as well as temperature on the mass modifications of the charmonium states
in the hadronic medium are small and the dominant medium effect
is observed to be the effect of density, in the present investigation. 

As already mentioned, in obtaining the mass shifts of the charmonium 
states as discussed above, we have taken into account the finite quark 
mass term in the trace of the energy momentum tensor and hence 
in the evaluation of the gluon condensate in the medium. 
However, if we do not consider the finite quark mass term in the
trace of the energy momentum tensor and calculate the mass shift 
of the charmonium states, then in the  isospin symmetric medium, 
at nuclear saturation density, 
$\rho_{B} = \rho_{0}$ and strangeness fraction $f_{s} = 0(0.5)$, 
the values of the mass shifts for the charmonium states $J/\psi$, 
$\psi(3686)$ and $\psi(3770)$ are observed to be $-9.32(-9.07)$,
$-126.4(-123)$ and $-167.5(-163)$ MeV respectively, for $N_f$=4 in the
expression of the one loop QCD beta function given by equation 
(\ref{qcdbeta}) and $-8.63(-8.41)$, $-117(-114)$ and $-155(-151)$,
when we take $N_f$=3 in the beta function, to obtain the expression
for the gluon condensate. For the limit of massless quarks,
the difference in the mass shifts of the charmonium masses
arises due to the different multiplying factors in the equation 
relating the scalar gluon condensate and the dilaton field. 
There is seen to be a larger mass shift for $N_f$=4,
due to the factor (24/25) in equation (\ref{chiglu}), 
which is larger than the factor (8/9) for the case of $N_f$=3.
The values of the mass shifts of these charmonium states 
in isospin symmetric matter at zero temperature for $\rho_B=\rho_0$
in the limit of massless quarks obtained for $N_f$=4, may be compared 
to the values of $-4.35 (-2.96)$, $-59 (-40)$ and $-78.5 (-54)$ MeV, 
when the finite quark mass term in the trace of the energy momentum 
tensor is taken into account. 
The effects of the finite quark mass on the masses of $J/\psi$ and 
$\eta_c$ obtained using the QCD sum rule approach 
due to the medium modifications of the gluon condensates
within the present chiral effective model, 
has been considered in Ref. \cite{charmmass2}. 
In Ref. \cite{charmmass2}, the value of the mass of $J/\psi$,
in symmetric nuclear matter at zero temperature for $\rho_B=\rho_0$, 
was obtained to be $-4.48(-8.01)$ MeV, with (without) accounting 
for the finite quark masses in the trace of the energy momentum tensor,
which may be compared with the value of mass shift of $J/\psi$
as $-4.35$ ($-9.32$) MeV in the present investigation. 
In the limit of massless quarks, the values of mass shifts of the 
charmonium states $J/\psi$, $\psi(3686)$ and $\psi(3770)$,
are observed to be $-34.72(-39.82)$, $-471(-540.5)$ and $-623(-714)$ 
MeV respectively, in symmetric hadronic matter 
for $f_{s} = 0(0.5)$, for baryon density $4\rho_{0}$.
In isospin asymmetric medium ($\eta$ = 0.5), at $\rho_{B} = \rho_{0}$
and $f_{s} = 0(0.5)$, these values are observed 
to be $-9.02(-9.37)$, $-122.5(-127)$ and $-162(-168)$ MeV.
Thus we observe that the magnitudes of the mass shifts
for the charmonium states are smaller when we consider the 
contribution of finite quark mass term for the modification
of the gluon condensate in the medium. The medium modification
of the gluon condensate is given in terms of the medium 
modifications of the dilaton field and the scalar fields, $\sigma$, 
$\zeta$ and $\zeta_c$, as can be seen from equation (\ref{gludiff}). 
As we have already mentioned, the fluctuation of the charm
scalar field $\zeta_c$ has been neglected in the present 
investigation for the study of the medium modifications of the
$D$, $\bar D$ mesons as well as for the study of the in-medium masses
of the charmonium states. The contribution of terms containing 
the scalar fields $\sigma$ and $\zeta$ arising from the finite quark 
mass term in the trace of the energy momentum tensor to the medium
modification of the gluon condensate given by equation (\ref{gludiff}),
are opposite in sign as compared to the first term given in terms
of the scalar dilaton field and this leads to the charmonium mass
shifts to be smaller when the finite quark mass term is accounted
for in the trace anomaly as compared to when it is neglected.
  
In Ref. \cite{leeko}, the mass modifications of the charmonium states 
were calculated in the symmetric nuclear medium at zero temperatures 
using QCD second order Stark effect in the linear density approximation. 
The mass shifts for the charmonium states $J/\psi$, $\psi(3686)$ 
and $\psi(3770)$ were found to be  $-8, -100$ and $-140$ MeV respectively
at the nuclear matter saturation density.
These may be compared to the values of -4.35, -59 and -78.5 MeV 
of the charmonium states at $\rho_B=\rho_0$ of
our present calculations when the effects of the finite quark masses
are taken into account in evaluation of the medium modification of
the gluon condensate and the values of -9.32, -126.4 and -167.5 MeV,
in the limit of massless quarks in the trace of energy momentum tensor
in QCD. In the present work, we have studied the isospin asymmetry and 
strangeness dependence of the masses of the charmonium states in the 
hadronic medium at finite temperatures, which will be relevant for 
the asymmetric heavy ion collision experiments planned at the future 
facility at GSI.
Using QCD sum rules and operator product expansion 
up to mass dimension four  \cite{klingl} and six \cite{kimlee}, 
the mass shift for $J/\psi$ mesons at nuclear saturation density 
was observed to be about $-7$ MeV and -4 MeV respectively. 
The effect of temperature on the $J/\psi$ in deconfinement phase was 
studied in \cite{leetemp,cesa} and it was reported 
that $J/\psi$ mass remains essentially constant within a wide range of 
temperature and above a particular value of the temperature, T, 
there is seen to be a sharp change in the mass of $J/\psi$ 
in the deconfined phase. For example, in Ref. \cite{lee3} 
the mass shift for $J/\psi$ was reported to be about 
200 MeV at T = 1.05 T$_{c}$. In the present work, we have studied the 
effects of temperature, density, isospin asymmetry and strangeness 
fraction, on the mass modifications of the charmonium states 
($J/\psi$, $\psi(3686)$ and $\psi(3770)$) in the confined hadronic phase, 
arising due to modifications of a scalar dilaton field 
which simulates the gluon condensates of QCD, as well as due
to fluctuations of the scalar fields $\sigma$ and $\zeta$ 
in the explicit chiral symmetry breaking term, within the 
chiral effective model, when the finite quark masses are taken 
into account in the trace of the energy momentum tensor of QCD. 
The effect of temperature is found to be 
small for the charmonium states $J/\psi$, $\psi(3686)$ and 
$\psi(3770)$, whereas the masses of charmonium states are observed to 
vary considerably with density, in the present investigation.
In Ref. \cite{leeko} the masses of the charmonium states have been 
studied due to the effects of the $D\bar D$ loop 
as well as of from the gluon condensates in the medium and 
the mass shifts for $J/\psi$, 
$\psi(3686)$ and $\psi(3770)$ mesons were observed to be $-5, 
-130$ and -125 MeV at nuclear saturation density in symmetric cold
nuclear medium, with the  medium modification of the charmonium masses 
arising from the modification of the gluon condensate dominating over
the contribution from the D meson loop. On the other hand, 
considering the effects of $D\bar D^*$, $D^* \bar D$ and 
$D^{*}\bar {D^*}$ meson loops, in addition to the $D\bar D$ loop, 
with the medium modifications of the D and $D^*$ mesons
calculated in a Quark meson coupling model, 
the mass shift for $J/\psi$ mesons at nuclear matter density 
was observed to range from -16 MeV to -24 MeV \cite{krein1},
with the contribution from the $D^*\bar D^*$ loop dominating over
the other terms. The range of the mass shifts of $J/\psi$ was
obtained depending on the value of the ultra-violet cut-off 
in the form factors used to regularize the loop integrals.

The strength of the harmonic oscillator wave function, $\beta$
can be modified in the hadronic matter, which can give an estimate
of the change in the size of the charmonium state in the medium. 
The shift in $\beta^2$ can be obtained by relating to the mass shift of the 
charmonium state given as $\delta M_{J/\psi}=\frac{3}{2M}\delta \beta^2$ 
for $J/\psi$ \cite{friman} and $\delta M_{\psi}=\frac{7}{2M}\delta \beta^2$ 
for the charmonium states $\psi(3686)$ and $\psi(3770)$.
We compute the root mean squared radii of the charmonium states 
from the mass shifts 
obtained in the present work, at given densities, and show the
obtained results in Fig. \ref{rmscharm} for the isospin symmetric 
($\eta$=0) and asymmetric (with $\eta$=0.5) strange hadronic matter.
These results are compared to the case of nuclear matter ($f_s$=0),
shown as dotted lines. We observe that the change in the radius of the
charmonium state $J/\psi$ from the vacuum value of 0.47 fms is marginal, 
even for high densities. For example,the radius of $J/\psi$
is modified to about 0.484 fms for $\rho_B=6\rho_0$. 
On the other hand, it is observed that
the sizes of the charmonium states $\psi(3686)$ and $\psi(3770)$, 
are seen to increase substantially at higher densities 
as can be seen in Fig. \ref{rmscharm}.
This is related to the fact that the mass shifts for these
mesons are large at high densities in the medium.
The value of the rms radii for  $\psi(3770)$ is seen to rise
faster than that of the $\psi(3686)$ at high densities.

The medium modifications of the masses of $D$ and $\bar {D}$ mesons
can have relevance to the experimental observable of $J/\psi$
suppression in relativistic heavy ion collision experiments. 
Due to the drop in the mass of the $D\bar D$ pair 
in the hadronic medium, it can become a possibility that the 
excited states of charmonium $\psi(3686)$ and $\psi(3770)$ can 
decay to $D\bar{D}$ pairs \cite{amarind} and hence the production 
of $J/\psi$ from the decay of these excited states can be suppressed. 
Even at some densities it can become a possibility that the $J/\psi$ 
itself decays to $D\bar{D}$ pairs. The effects of the medium modifications 
of the $D(\bar D)$ mesons on the decay widths for the decay of the 
charmonium states to $D\bar D$ pairs has been studied accounting 
for the internal structure of these mesons using the 3P0 model 
\cite{friman}. In this model, it was observed that the decay width 
does not increase monotonically with drop in the masses 
of the $D(\bar D)$ mesons with density as one would naively expect. 
On the contrary, the decay widths after an increase initially with 
decrease in the D($\bar D)$ meson masses was seen to decrease
with further drop in these masses and nodes were observed in 
the decay widths of the charmonium states $\psi(3686)$
and $\psi(3770)$ \cite{friman}.
In the present investigation, we study the effects of
the mass modifications of the $D(\bar D)$ mesons 
as well as of the charmonium states in the isospin
asymmetric hyperonic medium on the partial decay widths
of the charmonium states to $D^+D^-$ as well as $D^0\bar {D^0}$
pairs. We also observe nodes in the partial decay widths of
the charmonium states $\psi(3686)$ and $\psi(3770)$,
as has already been observed in the literature \cite{friman}, 
when the mass modifications of the $D(\bar D)$ mesons are taken
into account, but the mass modifications of the charmonium 
masses are neglected. However, there are observed to be
significant modifications to these partial decay widths,
when the mass modifications of the charmonium states
are also taken into account. There are no nodes observed
even up to a density of about 6$\rho_0$,
when the in-medium masses of the charmonium states are
taken into account in addition to the in-medium
$D$ and $\bar D$ meson masses.

\begin{figure}
\includegraphics[width=16cm,height=16cm]{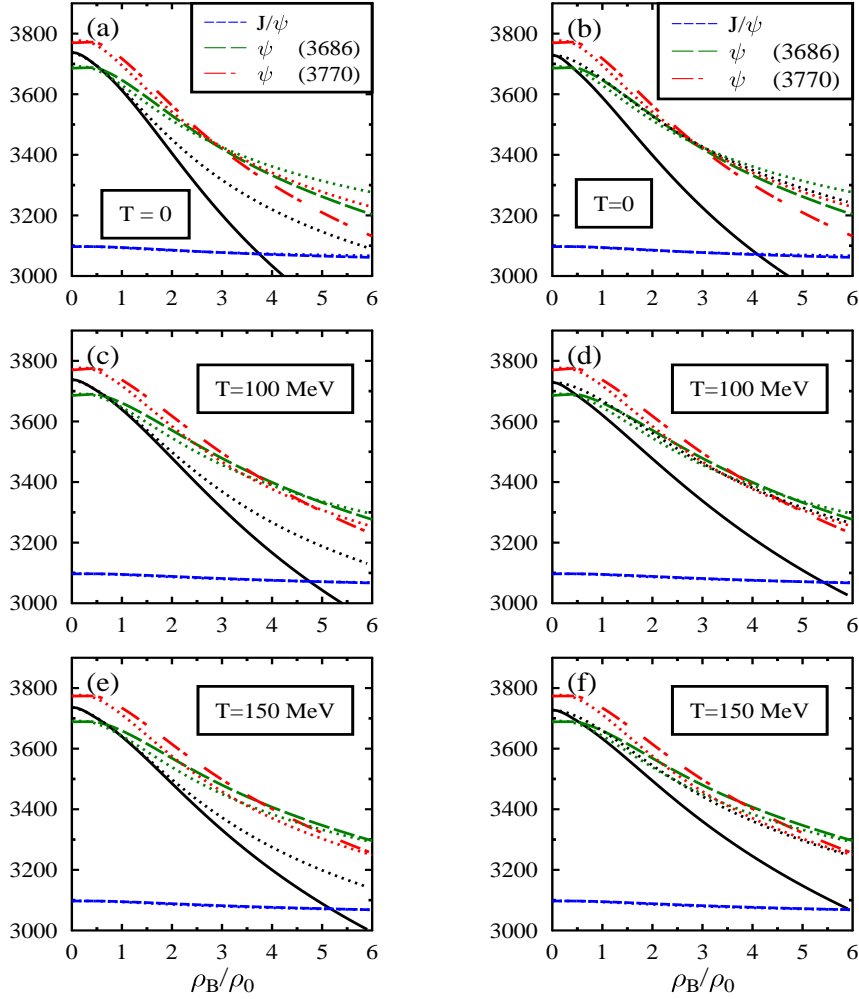}
\caption{(Color online) The masses of the charmonium states 
as well as the $D\bar{D}$ pair (shown by the solid line) 
[$D^{+}D^{-}$ in (a), (c), (e) and $D^{0}\bar{D}^{0}$ 
in (b), (d), (f)] in MeV plotted as functions of $\rho_{B}/\rho_{0}$ 
for isospin asymmetric strange hadronic matter ($\eta=0.5, f_{s}=0.5$) and 
for temperatures, T = $0, 100, 150$ MeV. The charmonium masses
have been calculated from the gluon condensate in the medium,
obtained by accounting for the finite masses of the quarks.
These results have been compared with the case of $f_s$=0
shown as dotted lines.} 
\label{jpsisuppmq}
\end{figure}

\begin{figure}
\includegraphics[width=16cm,height=16cm]{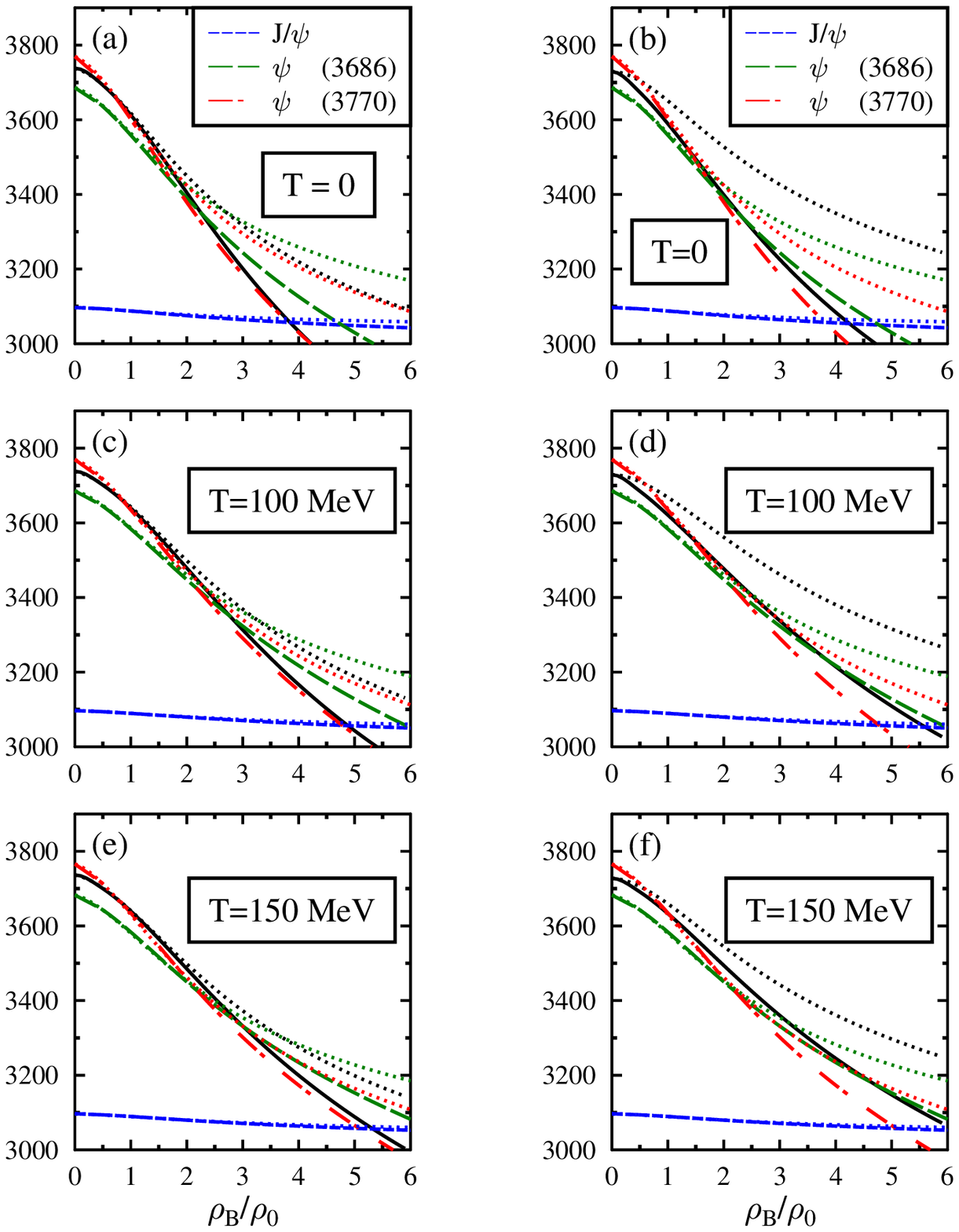}
\caption{(Color online) The masses of the charmonium states 
as well as the $D\bar{D}$ pair (shown by the solid line) 
[$D^{+}D^{-}$ in (a), (c), (e) and $D^{0}\bar{D}^{0}$ 
in (b), (d), (f)] in MeV plotted as functions of $\rho_{B}/\rho_{0}$ 
for isospin asymmetric strange hadronic matter ($\eta=0.5, f_{s}=0.5$) and 
for temperatures, T = $0, 100, 150$ MeV. The charmonium masses
have been calculated from the gluon condensate in the medium,
obtained by neglecting the finite masses of the quarks
in the trace anomaly. These results have been compared with
the case of $f_s$=0, shown as dotted lines.} 
\label{jpsisuppmq0}
\end{figure}

\begin{figure}
\includegraphics[width=16cm,height=16cm]{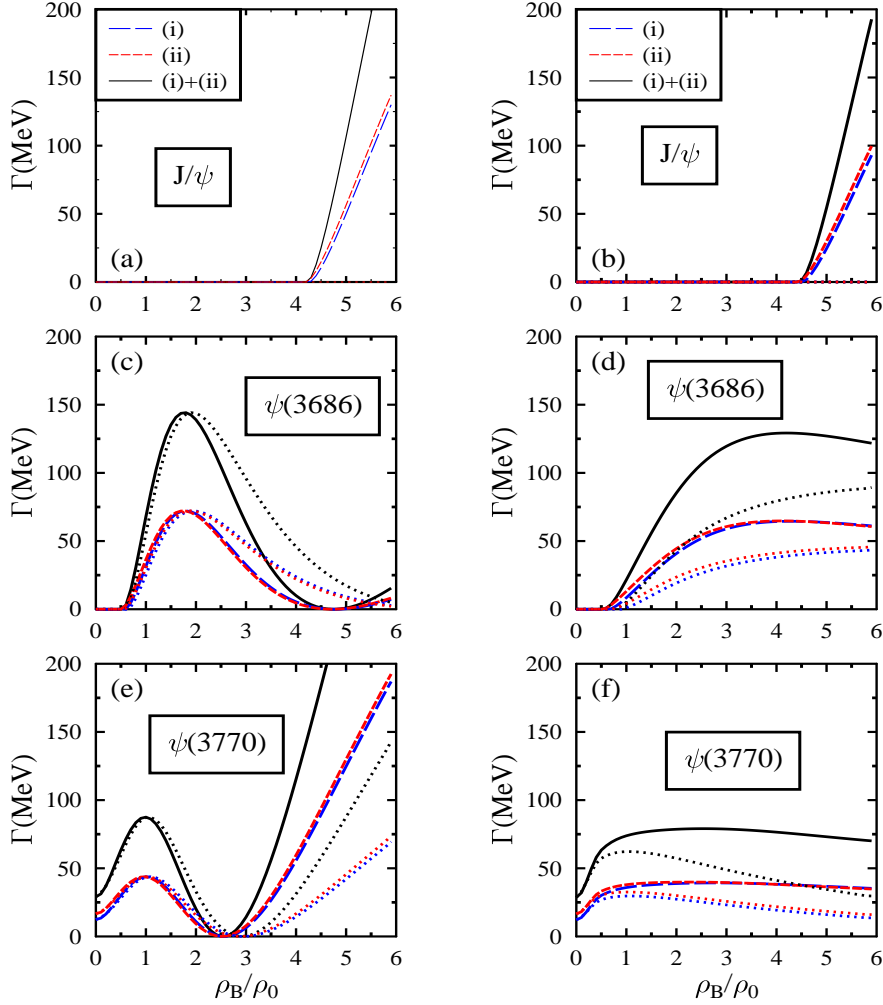}
\caption{(Color online) The partial decay widths of the charmonium states 
to (i) $D^+D^-$, (ii) $D^0\bar {D^0}$ and (iii) the sum of the 
two channels ((i)+(ii)) in the isospin symmetric strange hadronic matter 
($\eta$=0,$f_s$=0.5), 
accounting for the medium modifications of the $D(\bar D)$ mesons. 
These are shown in subplots (a), (c) and (e), when the mass modifications 
of the charmonium states are neglected and (b), (d) and (f), 
the partial decay widths are shown when the mass modifications 
of the charmonium states are also taken into account. These results
are compared to the case of $f_s$=0, shown as dotted lines.}
\label{charmdweta0}
\end{figure}

\begin{figure}
\includegraphics[width=16cm,height=16cm]{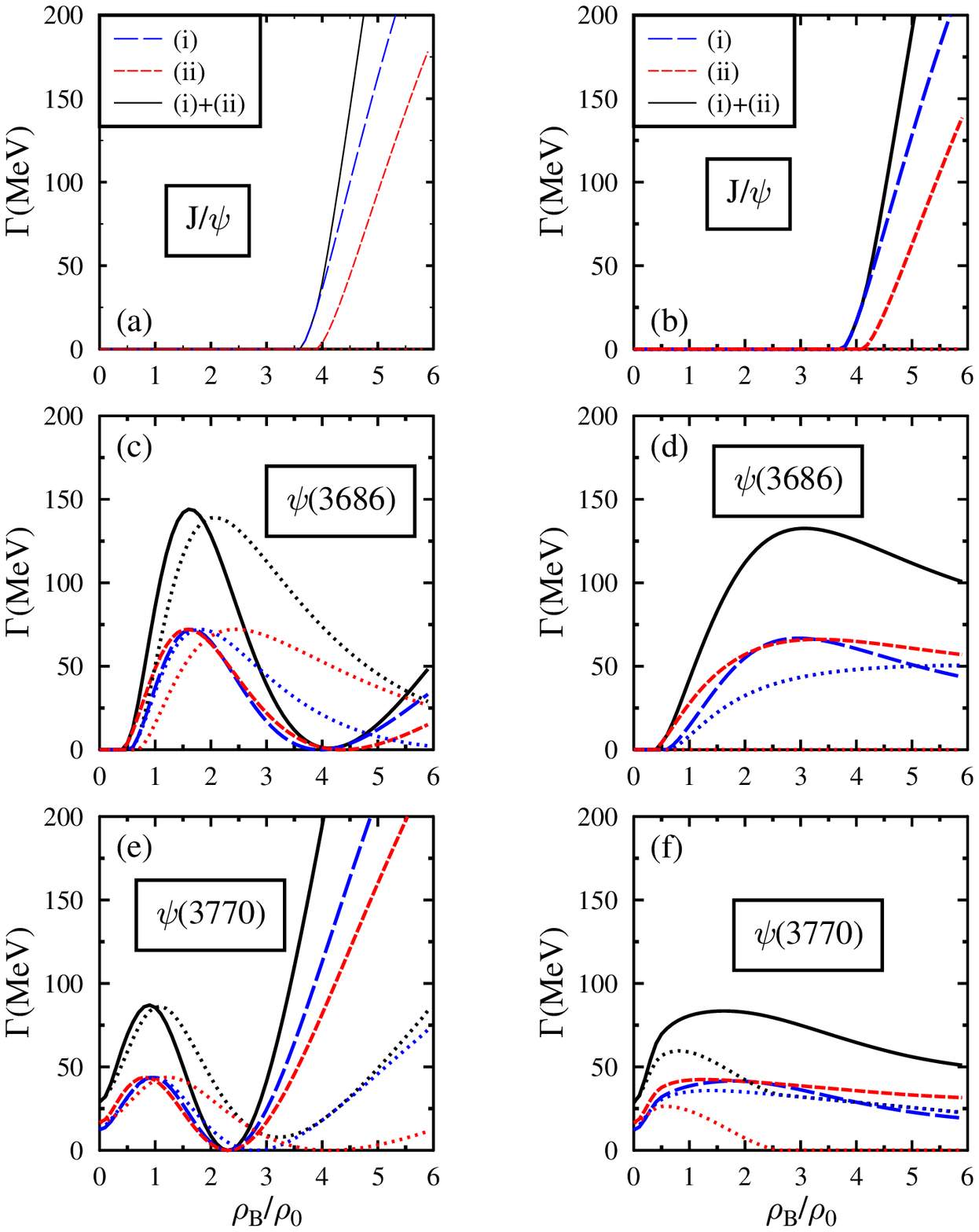}
\caption{(Color online) The partial decay widths of the charmonium states 
to (i) $D^+D^-$, (ii) $D^0\bar {D^0}$ and (iii) the sum of the 
two channels ((i)+(ii)) in the isospin asymmetric strange hadronic matter 
($\eta$=0.5, $f_s$=0.5) at T=0, as functions of the baryon density,
in units of the nuclear matter saturation density, 
accounting for the medium modifications of the $D(\bar D)$ mesons. 
These are shown in subplots (a), (c) and (e), when the mass modifications 
of the charmonium states are neglected and (b), (d) and (f), 
the partial decay widths are shown when the mass modifications 
of the charmonium states are also taken into account. These results
are compared to the case of $f_s$=0, shown as dotted lines.}
\label{charmdweta5}
\end{figure}

We next compare the in-medium masses of the charmonium states 
and the $D\bar D$ pair to investigate the possibility of the 
decay of the charmonium states to $D\bar D$ in the hadronic 
medium. In Fig. \ref{jpsisuppmq}, we show the density dependence of 
the masses of the $D^{+}D^{-}$ as well as $D^{0}\bar{D^{0}}$ pairs 
at strangeness fraction, $f_{s} = 0.5$ and for isospin asymmetry 
parameter, $\eta = 0.5$. These are shown for values of temperatures, 
T = 0, 100 and 150 MeV. In the same figure, we also show the 
in-medium masses of the charmonium states $J/\psi$, $\psi(3686)$ 
and $\psi(3770)$ considering the contribution of finite quark mass 
term in the trace of the energy momentum tensor in QCD.
We compare these results with the case of $f_s$=0 shown as
dotted lines. The masses of $D$ and $\bar D$ mesons are 
observed to decrease with inclusion of hyperons in the medium,
as has already been observed in Figs. \ref{dmass}
and \ref{dbarmass}, and hence with increase in $f_s$,
there is a drop in the mass of the $D\bar D$ pair.
However, there is seen to be a larger drop of the mass
of $D^0 \bar {D^0}$ as compared to $D^+D^-$ pair in the isospin
asymmetric medium. The masses of the charmonium states are also
observed to decrease with increase in $f_s$.
The possibility of the decay of charmonium states
to $D^+D^-$ as different from the decay to $D^0\bar {D^0}$ pair
in the isospin asymmetric medium,
is thus due to the difference in the mass modifications
of the $D^+D^-$ and $D^0\bar {D^0}$ pairs in the hadronic medium. 
We observe from Fig. \ref{jpsisuppmq}
that the in-medium mass of the excited charmonium state, $\psi(3770)$
always remains larger than the in-medium masses of $D^+D^-$
pairs in the asymmetric hadronic matter ($\eta$=0.5) for both 
$f_s$=0 and 0.5 and hence it can decay to $D^+D^-$ pairs
at all values of densities. However, it is observed that
the mass of $\psi(3770)$ remains larger than $D^0\bar {D^0}$ pair
for densities smaller than a density of around 2.8 $\rho_0$ 
for $\eta$=0.5 and $f_s$=0, 
and hence the decay of $\psi(3770)$ to $D^0\bar {D^0}$ is possible
only for densities less than this density. However,
for the asymmetric strange hadronic matter ($\eta$=0.5, $f_s$=0.5),
the decay of $\psi(3770)$ to $D^0\bar {D^0}$ is a possibility
for all densities. The mass of $\psi(3686)$ is observed to be 
larger than the in-medium mass of $D^+D^-$ pairs
above a baryon density of about $0.8 \rho_0 (0.5\rho_{0})$ for
$f_s$=0(0.5) and hence it is a possibility that this state can decay 
to $D^+D^-$ pairs above this value of density in the asymmetric
hadronic matter. The decay of these excited charmonium states 
to $D\bar D$ pairs may lead to $J/\psi$ suppression.
The $J/\psi$ mass is seen to be larger than the mass of the $D^+ D^-$ 
($D^0 \bar {D^0}$) pair above a density of 3.8 $\rho_0$ (4 $\rho_0$)
in the isospin asymmetric strange hadronic
matter ($\eta$=0.5, $f_s$=0.5) at zero temperature, which can lead to 
the decay of $J/\psi$ to $D\bar D$ pairs at high densities.
The density above which there can be possibility 
of the decay of $J/\psi$ to $D\bar D$ pairs is seen to increase  
as one increases the temperature of the hadronic medium. 
In Fig. \ref{jpsisuppmq}, we have investigated the medium modifications
of the charmonium masses in the strange hadronic matter, accounting
for the finite quark mass term in the trace of the energy momentum tensor
while evaluating the scalar gluon condensate in the medium. 
On the other hand, if we do not consider the contribution 
of finite quark masses to the in-medium properties 
of charmonium states, then there is seen to be a much larger drop
of the charmonium masses, which considerably modifies the conclusions
for the possibility of the decay of the charmonium states to $D\bar D$ 
pairs. We show these results in Fig. \ref{jpsisuppmq0}. For $\eta$=0.5
and $f_s$=0.5, the charmonium state $\psi(3770)$ is observed to have
a mass larger than the mass of the $D^+D^-$ ($D^0\bar {D^0}$) pair
for a density smaller than 0.8 $\rho_0$ (1.4 $\rho_0$), and hence the
decay to $D^+D^-(D^0 \bar {D^0})$ can be a possibility for densities 
smaller than this density. The charmonium state $\psi(3686)$
can decay to $D^+D^- (D^0\bar {D^0})$ for densities above
a density of around 2.2 $\rho_0$. For $J/\psi$, the decay to
$D^+D^- (D^0 \bar {D^0})$ seems possible above a density of
around 3.8 $\rho_0$ (4.2 $\rho_0$), when one does not account for
the quark mass term in the trace anomaly, for the computation of
the charmonium masses from the medium modification of the gluon
condensate.

The possibility of the decay of the charmonium states to the $D\bar D$
pairs has been investigated in isospin asymmetric nuclear matter 
at finite temperature within the model used in the present investigation 
\cite{amarvind}. However, in this work \cite{amarvind}, the effect 
of the finite quark masses were neglected in the trace of the energy 
momentum tensor for the evaluation of the gluon condensate modifying 
the masses of the charmonium states in the medium. As we have already 
discussed, the charmonium states are observed to have much smaller drop in  
the medium when the finite quark masses are taken into account
leading to a larger possibility of the decay of charmonium states
to the $D\bar D$ pairs. In Ref. \cite{amarvind}, due to the larger
drop in the charmonium masses in the limit of massless quarks, as 
compared to the present investigation, where the finite quark masses
are taken into consideration, the charmonium states decaying to
the $D\bar D$ pairs was becoming a possibility only for high 
densities, when the charmonium masses were larger than the mass
of the $D\bar D$ pair. The effect of strangeness on the masses
of the D and $\bar D$ mesons in the present investigation is
to decrease the masses of the $D$ and $\bar D$ mesons as well as
of the charmonium states in the massless quark limit. However,
while accounting for the finite quark mass term in the trace of
energy momentum tensor, the gluon condensate in the medium
has contributions from the dilaton field as well as from the 
scalar $\sigma$ and $\zeta$ fields, which is seen to 
the observed behavior of smaller drop of the charmonium 
states in the medium as compared to the case for massless quark 
limit, as shown in Fig. \ref{charmdelm}.

\subsection{Partial Decay widths of the charmonium states 
to $D\bar D$ pairs}

We compute the partial decay widths of the charmonium states
to $D\bar D$ pairs by accounting for the internal structure of these
mesons using the 3P0 model. These decay widths are calculated 
by using the expressions as given by equations (\ref{dwjpsi}),
(\ref{dwpsi3686}) and (\ref{dwpsi3770}), which are computed 
from the matrix element for the specific charmonium state 
decaying to $D\bar D$ pair. The wave functions for the charmonium 
state as well as the $D(\bar D)$ mesons are assumed to be 
of harmonic oscillator, with strengths $\beta$ and $\beta_D$ 
respectively. As has already been mentioned in the previous section, 
the value of the parameter $\beta$ is fitted from the rms radius 
of the specific charmonium state. The strength of the wave function
for the $D(\bar D)$ as well as the strength for the decay of
the charmonium state to $D(\bar D)$ are fitted from the
decay of $\psi(3770)$ to $D(\bar D)$ in vacuum. 
As one might observe from the expressions for the decay widths
of the charmonium states to $D(\bar D)$ given by equations 
(\ref{dwjpsi}), (\ref{dwpsi3686}) and (\ref{dwpsi3770}), 
the dependence of these decay widths
on the medium is through the magnitude of the center of mass 
momentum of the produced $D(\bar D)$ meson, $p_D$, written
as x in dimensionless units ($x=p_D/\beta_D$).
The value of $p_D$ depends on the medium through
the modifications of the masses of the charmonium state,
as well as on the masses of the $D\equiv (D^0,D^+)$ 
and $\bar D \equiv (\bar {D^0},D^-)$ mesons, as can be seen from
its expression given by (\ref{pd}).

In Figs. \ref{charmdweta0} and \ref{charmdweta5},
we show the partial decay widths of (i) $\psi \rightarrow D^+D^-$,
(ii) $\psi \rightarrow D^0 \bar {D^0}$ and (iii) $\psi \rightarrow 
D^+D^-$ and $\psi \rightarrow D^0\bar {D^0}$, for the charmonium state,
$\psi$ as $J/\psi$, $\psi(3686)$ and $\psi(3770)$ for zero temperature, 
for the symmetric ($\eta$=0) and asymmetric (with $\eta$=0.5) hadronic 
matter. These have been shown for the value of strangeness fraction, 
$f_s$=0.5 and have been compared with the case of nuclear matter 
($f_s$=0). For the case of symmetric nuclear matter ($\eta$=0 and 
$f_s$=0), the decay width of the charmonium state $\psi(3686)$ 
vanishes below a density of $\rho_B=0.6\rho_0$ as the mass of the
charmonium state is smaller than the mass of the $D(\bar D)$
pair ($D^+D^-$ as well as $D^0 \bar {D^0}$) for these densities. 
For densities above this value of the density, the magnitude of $p_D$ 
is seen to increase with density. However, at high densities, 
the increase is seen to be much slower when the density is increased 
still further. The polynomial part 
$\Big (1+\frac {2 r^2(1+r^2)}{(1+2r^2)(3+2r^2)(1-3r^2)}x^2\Big)^2$,
in the expression for the decay
width of $\psi(3686)$ given by equation (\ref{dwpsi3686})
remains unity for densities less than
$0.6 \rho_0$ (since $x=p_D/\beta_D$ vanishes 
for these densities) and is seen to decrease with 
density, for higher densities. 
The decay width for the decay of $\psi(3686) \rightarrow
D\bar D$, which is a combined effect of the polynomial as well as 
the gaussian parts, 
is seen to have an initial increase with density and 
then a drop as the density is further increased.
However, one does not observe any nodes in the partial
width even up to a density of about 6$\rho_0$.
With the inclusion of strangeness in the medium,
the behavior of the decay width of $\psi(3686)\rightarrow 
D\bar D$ with density is seen to be similar to as in the 
nuclear matter. However, due to smaller values of the
masses of the $D(\bar D)$ mesons, leading to higher values
of the center of mass momentum, $p_D$, there is faster suppression
of the polynomial part of the decay width leading to nodes
in the decay width of $\psi(3686)$ to $D^+D^-$ as well as to
$D^0\bar {D^0}$ at a density of about 4.5 $\rho_0$,
as can be seen in the subplot (c) in figure \ref{charmdweta0}.
Such nodes have been already observed in the literature 
in the decay widths of the charmonium states $\psi(3686)$ 
and $\psi(3770)$ to $D\bar D$ \cite{friman}, when the drop 
in the masses of the $D(\bar D)$ mesons was incorporated.
The decay of $\psi(3770) \rightarrow D\bar D$ is possible
in vacuum as the mass of $\psi(3770)$ is higher than the mass 
of the $D\bar D$ pair. For $\psi(3770)$, there is seen to be 
initial increase in $p_D$ as a function of density and then 
a drop with further increase in the density when the modifications 
of the $D(\bar D)$ masses as computed in the present investigation,
are taken into account. However, there is seen to a strong suppression 
of the decay width with density due to the polynomial part,
$\Big (1-\frac {1+r^2}{5(1+2r^2)}x^2\Big)^2$
of the expression for the decay width given in
equation (\ref{dwpsi3770}). Consequently, 
in the present investigation, there are seen to be nodes 
in the partial decay widths of the charmonium states $\psi(3686)$ 
and $\psi(3770)$ to $D\bar D$ pairs at densities of about 4.5$\rho_0$
and 2.8 $\rho_0$ respectively (see subplots (c) and 
(e) of figure \ref{charmdweta0}) in symmetric strange hadronic 
matter, when the mass modifications
of the $D$ and $\bar D$ mesons are taken into account, but
the modifications of the charmonium masses are neglected.
One observes the nodes in the decay widths for these charmonium
states also for the asymmetric hadronic matter, as can be seen
from subplots of (c) and (e) of figure \ref{charmdweta5}. 
However, the nodes in the decay widths are observed at 
smaller values as compared to the isospin symmetric case,
since the value of $p_D$ increases with isospin asymmetry
in the medium, due to larger drop in the $D(\bar D)$ masses.
On the other hand, there are seen to be significant modifications
to the partial decay widths of $\psi(3686)$ and $\psi(3770)$,
when the mass modifications of these charmonium states 
are also taken into account, as can be seen from
the subplots (d) and (f) in figures \ref{charmdweta0}
and \ref{charmdweta5}. This is due to the fact that
the value of $p_D$ decreases for the latter case,
due to the drop in the charmonium masses in the medium,
as can be seen from the expression for $p_D$ given by equation
(\ref{pd}). The behavior of the decay widths of an initial 
rise and then a drop as we increase the density, is still
observed, when the mass modifications of the charmonium states
as well as $D$ and $\bar D$ are taken into account.
However, the suppression of the decay width is not as large
as in the case of neglecting the charmonium mass 
(due to the smaller value of $p_D$ for a given density)
and consequently, no nodes are observed even up to a density 
of about 6$\rho_0$. For the case of $\eta$=0.5, as can be seen 
from the subplot (f) of figure \ref{charmdweta5}, the decay 
of the state $\psi(3770)$ to $D^+D^-$ is a possibility 
for all densities, whereas its decay to $D^0\bar {D^0}$ 
does not become possible above a density of about 2.8$\rho_0$ 
for $f_s$=0, as can also be seen from subplots (a) and (b) 
of figure \ref{jpsisuppmq}. For $J/\psi$, for densities above 
the threshold density for the decay of $J/\psi$ to $D\bar D$ 
pairs, the decay width is seen to increase monotonically with 
density, and there is no polynomial part in the the expression 
of the decay width of $J/\psi$ (see equation (\ref{dwjpsi})) 
and so no nodes are observed in the decay width of $J/\psi$ 
to $D\bar D$ pair.

\section{Summary}
To summarize, in the present investigation, we have studied the in-medium
properties of $D$ and $\bar{D}$ mesons in isospin asymmetric strange 
hadronic matter at finite temperatures by generalizing the chiral 
SU(3) model to SU(4). The properties of $D$ mesons are modified 
due to interaction with the baryons (nucleons and hyperons) and 
also with the scalar fields $\sigma$, $\zeta$, $\delta$ and 
the dilaton field $\chi$. The scalar meson-baryon coupling parameters 
of the model are fitted from the vacuum hadron masses and the parameters 
$d_1$ and $d_2$ occurring in the range terms are fitted from the low 
energy $KN$ scattering data. We observe that for a given value of 
density and isospin asymmetry, the strangeness of the medium is seen 
to lead to decrease in the masses of $D$ and $\bar{D}$ mesons and 
the masses of the $\bar D$ mesons are seen to be more sensitive 
to the strangeness of the medium as compared to the masses of the 
$D$ mesons. On the other hand, in the isospin asymmetric strange 
hadronic matter, the masses of the $D$ mesons are observed 
to depend more strongly on the isospin asymmetry of the medium,
as compared to the masses of the $\bar{D}$ mesons. 
The mass shifts for the $D$ mesons obtained in the present calculations 
are observed to be appreciable and the values 
are in agreement with the calculations of the mass shifts
in the QMC model \cite{qmc} as well as in QCD sum rule approach 
\cite{qcdsum08,weise}. On the other hand, the calculations with the
coupled channel approach \cite{ljhs,mizutani8} observed only small
mass modifications of the $D$ and $\bar{D}$ mesons in the medium.
The $D$ mesons are also observed to have attractive mass shifts
within the calculations using a coupled channel approach based 
on heavy quark symmetry \cite{garcia1}. However, the mass shifts
in Ref. \cite{garcia1} are observed to be much smaller than
what is observed in the present work. The attractive mass shifts
of the $D$ mesons can give rise to the possibility of the $D$-mesic 
nuclei, which have been investigated in Ref. \cite{garcia1}. 
The ratios $D^{+}/D^{0}$ and $D^{-}/\bar{D^{0}}$ of their
production cross sections as well as collective flow, could be 
promising observables to study the effect of strangeness fraction 
of the medium on the properties of $D$ and $\bar{D}$ mesons.
The isospin dependence of $D^{+}$  and $D^{0}$ masses is seen to be 
a dominant medium effect at high densities, whereas, for the $D^{-}$ 
and $\bar{D}^{0}$, one sees that, even though these have a strong 
density dependence, their in-medium masses remain similar at a given 
value for the isospin-asymmetry parameter $\eta$. As the production 
of the $D$ and $\bar D$ mesons are dominantly by the decay of the 
charmonium states to $D^+D^-$ ($D^0\bar {D^0}$) pairs, the production 
cross-sections for the ratios of $D^+/D^0$ as well as $D^-/\bar {D^0}$ 
are expected to be modified substantially in the medium due to the 
stronger isospin dependence of the $D$ meson doublet as compared 
to the $\bar D$ meson doublet, whereas, the parameter, $v_2$ 
for the anisotropic collective flow, is expected 
to modify the ratio of $D^+/D^0$ more appreciably as compared to
the flow of the ratio $D^-/\bar {D^0}$. However, due to the presence
of the light antiquark (quark) in the $D(\bar D)$, they might undergo
substantial decay in the hadronic environment, which could make it
difficult to extract clean signals for the medium modifications
of the $D$ and $\bar D$ mesons in the isospin asymmetric
strange hadronic medium, from the experimental data arising
from isospin asymmetric heavy ion collision experiments.

We have also investigated in the present work, the effects of density, 
temperature, isospin asymmetry and strangeness fraction 
of the strange hadronic medium on the masses 
of the charmonium states $J/\psi$, $\psi(3686)$ and $\psi(3770)$, 
arising due to modification of the scalar dilaton field, $\chi$, 
which simulates the gluon condensates of QCD, within the chiral 
SU(3) model, as well as due to the modifications of the scalar fields
$\sigma$ and $\zeta$, when we account for the quark mass term in
the trace of the energy momentum tensor in QCD. We have taken 
into consideration the finite quark 
masses through the explicit chiral symmetry breaking term in the chiral
effective model, for estimating the medium modification
of the gluon condensates, from which we obtain the mass
modifications of the charmonium states in the present investigation. 
Due to the modifications of the charmonium wave functions
in the hadronic medium, the radii of the charmonium states
are modified and these have been computed in the present work.
For the charmonium states $\psi(3686)$ and $\psi(3770)$,
there is seen to be increase in their root mean square radii,
with density, arising due to the decrease in the strengths of their
harmonic oscillator wave functions in the hadronic medium.
The change in these wave functions are computed from the
mass shifts of these charmonium states as calculated 
in the present investigation.

The partial decay widths for the decay of the charmonium states
to $D^+D^-$ as well as $D^0\bar {D^0}$ have also been investigated
in the hadronic medium, which are seen to have nodes 
for the charmonium states $\psi(3686)$ and $\psi(3770)$, 
as has already been observed in the literature \cite{friman},
when we account for the mass modifications of the D and $\bar D$ mesons, 
but do not consider the in-medium masses of the charmonium states.
However, while accounting for the mass modifications of the
charmonium states as well, the partial decay widths 
are seen to be modified significantly, and there are no nodes
in the partial decay widths observed up to a density of about
6$\rho_0$.
The change in the mass of $J/\psi$ with the density 
is observed to be small at nuclear matter saturation density and 
is in agreement with the QCD sum rule calculations. There is seen 
to be appreciable drop in the in-medium masses of excited charmonium 
states $\psi(3686)$ and $\psi(3770)$ with density. The mass drop 
of the excited charmonium states $\psi(3686)$ and  $\psi(3770)$ 
are large enough to be seen in the dilepton spectra emitted from 
their decays in experiments involving $\bar{p}$-A annihilation 
in the future facility at GSI, provided these states decay 
inside the nucleus. The in-medium properties of the excited 
charmonium states $\psi(3686)$ and $\psi(3770)$ can be studied 
in the dilepton spectra in $\bar{p}$-A experiments in the future 
facility of the FAIR, GSI \cite{gsi}. The mass shift of the charmonium 
states in the hot hadronic medium seem to be appreciable at high densities 
as compared to the temperature effects on these masses, and these should 
show in observables like the production of these charmonium states in the
compressed baryonic matter experiment at the future facility at GSI, 
where baryonic matter at high densities and moderate temperatures will
be produced.

The medium modifications of the $D$ meson masses can lead to a suppression 
in the $J/\psi$ yield in heavy-ion collisions, since the excited states of 
the $J/\psi$, as well as $J/\psi$, can decay to $D\bar{D}$ pairs 
in the dense hadronic medium. The medium modifications of the masses of
the charmonium states as well as the $D$ and $\bar{D}$ mesons have been
considered in the present investigation in strange 
hadronic medium at finite temperatures. In the present investigation,
the charmonium masses have been calculated taking into account
the finite quark mass term in the trace of the energy momentum tensor
and the mass drop for the charmonium state is seen to be appreciably 
smaller than when this is not taken into account. We observe that
the excited charmonium states, $\psi(3686)$ as well as $\psi(3770)$
have a possibility of decaying to $D\bar D$ pairs, whereas $J/\psi$
is observed to decay to $D\bar D$ pair only above certain value
of the density depending on the isospin asymmetry and temperature
of the medium. The conclusions for the possibility of the decay
of the charmonium states to the $D\bar D$ pairs are found to be 
substantially modified when the in-medium masses of the charmonium 
states are calculated from the gluon condensate in the medium, obtained
by neglecting the quark mass term in the energy momentum tensor in QCD. 
The strong density, strangeness and isospin dependence of the $D(\bar{D})$ 
meson masses, charmonium masses as well as the partial decay widths
of the charmonium states decaying to $D\bar D$ pairs, in hot isospin 
asymmetric strange hadronic matter can be tested in the asymmetric 
heavy-ion collision experiments at future GSI facility \cite{gsi}. 

We thank Arshdeep Singh Mehta for discussions. Financial support from 
Department of Science and Technology, Government 
of India (project no. SR/S2/HEP-21/2006) is gratefully acknowledged. 
One of the authors (AM) is grateful to the Frankfurt Institute for Advanced
Research (FIAS), University of Frankfurt, for warm hospitality and 
acknowledges financial support from Alexander von Humboldt Stiftung 
when this work was initiated.

\end{document}